\documentclass[lettersize,journal]{IEEEtran}
\usepackage{amsmath,amsfonts}
\usepackage{algorithmic}
\usepackage{array}
\usepackage[caption=false,font=normalsize,labelfont=sf,textfont=sf]{subfig}
\usepackage{textcomp}
\usepackage{stfloats}
\usepackage{url}
\usepackage{verbatim}
\usepackage{graphicx}
\usepackage{cite}
\usepackage{amsmath}
\usepackage{amssymb}
\usepackage{bm} 
\usepackage[ruled,linesnumbered,noline]{algorithm2e}
\def\linspread{0.95}
\begin{document}

\title{Cross-Sparsity-Enabled Multipath Perception via Structured Bayesian Inference for Multi-Target Estimation}

\author{Xiang Chen,~\IEEEmembership{Student Member,~IEEE,}~Ming-Min Zhao,~\IEEEmembership{Senior Member,~IEEE,}~An Liu,~\IEEEmembership{Senior Member,~IEEE,} Min Li,~\IEEEmembership{Senior Member,~IEEE,}~Qingjiang Shi,~\IEEEmembership{Senior Member,~IEEE,} and Min-Jian Zhao,~\IEEEmembership{Member,~IEEE}
        % <-this % stops a space
%\thanks{This paper was produced by the IEEE Publication Technology Group. They are in Piscataway, NJ.}% <-this % stops a space
\thanks{Xiang Chen, Ming-Min Zhao, An Liu, Min Li, and Min-Jian Zhao are with the College of Information Science and Electronic Engineering, Zhejiang University, Hangzhou 310027, China (e-mail: 12231089@zju.edu.cn; zmmblack@zju.edu.cn; anliu@zju.edu.cn; min.li@zju.edu.cn; mjzhao@zju.edu.cn).}
	\thanks{Qingjiang Shi is with the School of Computer Science and Technology, Tongji University, Shanghai 201804, China, and also with the Shenzhen Research Institute of Big Data, Shenzhen 518172, China (e-mail: shiqj@tongji.edu.cn).}
}



\maketitle

\begin{abstract}
In this paper, we investigate a multi-target sensing system in multipath environment, where inter-target scattering gives rise to first-order reflected paths whose angles of departure (AoDs) and angles of arrival (AoAs) coincide  with the direct-path angles of different targets. Unlike other multipath components, these first-order paths carry structural information that can be exploited as additional prior knowledge for target direction  estimation. To exploit this property, we construct a sparse representation of the multi-target sensing channel and propose a novel cross sparsity structure under a three-layer hierarchical structured (3LHS) prior model, which leverages the first-order paths to enhance the prior probability of the direct paths and thereby improve the estimation accuracy. Building on this model, we propose a structured fast turbo variational Bayesian inference (SF-TVBI) algorithm, which integrates an efficient message-passing strategy  to enable tractable probabilistic exchange within the cross sparsity, and a two-timescale update scheme to  reduce the update frequency of the high-dimensional sparse vector.  Simulation results demonstrate that leveraging the proposed cross sparsity structure  is able to improve the target angle estimation accuracy substantially, and  the SF-TVBI algorithm achieves estimation performance comparable to that of the Turbo-VBI, but with lower computational complexity.
\end{abstract}


\section{Introduction}
 Recently, the impact of multipath propagation has received considerable attention in sensing research, as it plays a critical role in practical environment perception \cite{8170154,6832777,6985926,8961111,8550811,9296833,5764515}. In such environments, radar sensing performance is often degraded because the received echoes comprise not only the desired line-of-sight (LoS) returns but also various multipath components \cite{richards2005fundamentals}. The presence of these multipath signals complicates target estimation and poses significant challenges for achieving reliable signal processing \cite{kamann_automotive_2018}.

This multipath effect occurs when the target’s echo takes multiple paths to reach the receiver, including both direct and indirect paths \cite{sabet_hybrid_2020}. Due to the fact that the angle of departure (AoD) and the angle of arrival (AoA) are not identical for certain indirect paths, the presence of such non-line-of-sight (NLoS) components may adversely affect the estimation accuracy of the direct-path angles~\cite{levy-israel_mcrb_2023}. To address this issue, one effective approach is to suppress the multipath effects~\cite{visentin_analysis_2017,li_fans-shaped_2021,manzoni_multipath_2023,zheng_detection_2024}. For example, an  AoA detection scheme based on coherent Pauli decomposition was proposed in~\cite{visentin_analysis_2017}, which enables the separation of objects with different polarimetric features that are located at the same radial distance from the radar. {Complementary to such signal-processing-based techniques}, a high-gain fan-shaped beam (FSB) antenna was designed in~\cite{li_fans-shaped_2021}, which enhances radar imaging performance by mitigating multipath effects induced by ground reflections. A vehicle-based multiple-input multiple-output (MIMO)-synthetic aperture radar (SAR)  imaging method was demonstrated in \cite{manzoni_multipath_2023} that can  effectively suppress multipath effects by taking an antenna-domain technique that synthesizes a two-dimensional (2D) aperture. Besides,  in \cite{zheng_detection_2024},  a generalized likelihood ratio test (GLRT) framework and a sparsity-enforced compressed sensing (CS) approach with Levenberg–Marquardt (LM) optimization were proposed to mitigate multipath effects.
However, the improvement in estimation accuracy achieved by suppressing multipath effects is inherently limited, because it forgoes path-diversity gains and cannot perfectly remove residual multipath in practice.

To address the above limitation, the potential of exploiting multipath effects for radar sensing performance enhancement  has attracted considerable attention in recent years, particularly in the context of target detection \cite{sen_ofdm_2010,sen_adaptive_2011,rong_diffuse_2020,xu_mimo_2021}.
Early efforts focused on waveform diversity, for instance, an information-theoretic waveform design was proposed in \cite{sen_ofdm_2010} to exploit multipath through orthogonal frequency division multiplexing (OFDM) frequency diversity and polarization, thereby enhancing radar detection and tracking.
Building upon this idea, parametric detection methods under the generalized multivariate variance  analysis framework were introduced in \cite{sen_adaptive_2011}, where an adaptive waveform design was shown to further improve detection performance.
More recently, robustness issues were considered and  an adaptive constrained generalized likelihood ratio test was proposed in \cite{rong_diffuse_2020}, which is composed of an expectation likelihood based uncertainty estimation and a Rao test,  both endowed with robustness against unknown disturbance covariance while maintaining a constant false alarm rate. To mitigate the effects of random sensing signal phase, a joint waveform-filter design was proposed in \cite{xu_mimo_2021} through a robust semi-definite relaxation approach.
However, the above studies mainly  focused on exploiting multipath effects to improve radar detection performance, while limited attention was paid to enhancing target estimation accuracy. Moreover, most of these works only focused on  performance improvement through signal-to-noise ratio optimization, without delving into the design and analysis of specific estimation algorithms.

In CS, structured sparsity  is a key mechanism for improving estimation performance \cite{bajwa_compressed_2010,potter_sparsity_2010,tang_compressed_2013,zheng_super-resolution_2017,chen_sparse_2018,chen_off-grid_2019,liu_robust_2020,xu_joint_2024,chen_joint_2024,zhou_robust_2025}. Specifically, the  notion of multipath sparsity in wireless channels was formalized in \cite{bajwa_compressed_2010}, where a compressed channel sensing (CCS) framework was proposed to leverage CS theory for efficient sparse channel estimation. In parallel, \cite{potter_sparsity_2010} introduced the use of sparse reconstruction algorithms and randomized measurement strategies for radar processing, demonstrating their potential to significantly enhance estimation performance. Besides, the challenge of gridless sparse estimation was addressed in  \cite{tang_compressed_2013} through the introduction of atomic norm minimization, which enables the recovery of frequency-sparse signals directly in the continuous domain with rigorous performance guarantees. This idea was further extended in \cite{zheng_super-resolution_2017} to the joint delay--Doppler domain for passive radar, where atomic-norm sparsity characterizes the continuous delay-Doppler structure and $\ell_{1}$-norm sparsity is imposed to capture demodulation errors, thus enriching the sparsity modeling paradigm. However, these works mainly rely on conventional CS algorithms for target/channel estimation, sparsity-aware priors are in general underutilized which may limit the potential gains in estimation accuracy.

To fully exploit the gains brought by structured sparsity,   structured-sparsity prior models  and  variational Bayesian inference (VBI) based sensing methods  were proposed in  \cite{liu_robust_2020,xu_joint_2024,chen_joint_2024}. Specifically, a three-layer hierarchical structured (3LHS) prior with a cluster-sparse Markov chain was introduced in \cite{liu_robust_2020}, and a flexible Bayesian framework was provided to capture diverse structured sparsities while maintaining tractability. Building upon this idea, a spatially nonstationary Markov random field (MRF) model was employed in \cite{xu_joint_2024} to characterize 2D burst sparsity in joint radar and communication channels, thereby exploiting the spatial correlations of scatterers. In \cite{chen_joint_2024}, partially overlapping structured (POS) sparsity and 2D block sparsity were jointly modeled, which further extends the scope of structured sparsity to incorporate overlapping and block-level dependencies inherent in sensing and communication (SAC) channels. However, these studies were limited to scenarios with single-bounce paths and idealized angle associations, 
and  did not account for first-order  multipath components (i.e., the propagation paths that are  reflected by one target and then by another before reaching the receiver).

Motivated by the above considerations, we investigate a multi-target sensing system in a multipath environment, as illustrated in Fig. \ref{fig8}. A key insight of this work is that the first-order multipath components, which have been shown to possess non-negligible energy and to play a critical role in target estimation \cite{levy-israel_mcrb_2023,zheng_detection_2024,9542970,9764274,10063228}, should not simply be considered as interference. Instead, they carry valuable structural information that can be explicitly exploited to improve target estimation performance. Specifically, these paths are characterized by the fact that their AoAs and AoDs coincide with the direct-path angles of different targets. We demonstrate how this distinctive property can be exploited as a powerful prior to significantly improve estimation accuracy. The main contributions of this paper are summarized as follows.
\begin{itemize}
	\item  
We construct a sparse representation of the multi-target sensing signal in a multipath scenario based on two-dimensional (2D) angular grids, where the transmit and receive angles are decoupled to effectively capture the presence of first-order paths. Furthermore, by exploiting the  property that the AoDs and AoAs of a first-order path respectively coincide with  the  direct angle of different  targets, we propose a novel \emph{cross sparsity} structure. This structure enables us to effectively  leverage the prior information conveyed by first-order paths to assist the estimation of direct paths, thereby enhancing the overall target estimation performance.
	\item 
A structured fast turbo variational Bayesian inference (SF-TVBI) algorithm is proposed that integrates VBI, message passing, and expectation–maximization (EM) to jointly obtain the marginal posteriors of the sparse multi-target sensing channel and the maximum likelihood (ML) estimates of the angular offsets. In particular, a two-timescale strategy and a grid-selection strategy are introduced to  reduce the number of E-step iterations and restrict the updates to only the most relevant grids, respectively. Meanwhile, to address the intractability of direct message passing caused by the numerous loops in the proposed factor graph, an efficient message-passing strategy is developed to enable tractable probabilistic information exchange. 
\item 
Simulation results validate the effectiveness of the proposed cross sparsity structure, and show that the proposed SF-TVBI algorithm strikes a favorable balance between accuracy and complexity. In particular, it surpasses the orthogonal matching pursuit (OMP) in performance, rivals turbo variational Bayesian inference (Turbo-VBI) in estimation accuracy, and yet maintains significantly lower computational cost through its two-timescale updates and grid-selection strategy.
\end{itemize}

The rest of this paper is organized as follows. In Section II, we present the system model for multi-target sensing in a multipath scenario. In Section III, we construct the sparse presentation of the multi-target sensing signal and introduce the proposed cross sparsity structure. Section IV proposes the SF-TVBI algorithm to estimate the target positions. Numerical results are provided in Section V and finally we conclude the paper in Section VI.

\emph{Notations}: Scalars, vectors and matrices are respectively denoted by lower/upper case, boldface lower case and boldface upper case letters. For an arbitrary matrix $\mathbf{A}$, $\mathbf{A}^T$, $\mathbf{A}^*$ and $\mathbf{A}^H$ denote its transpose, conjugate and conjugate transpose respectively. $\|\cdot\|$ denotes the Euclidean norm of a complex vector, and $\lvert\cdot\lvert$ denotes the absolute value of a complex scalar. $\lceil \cdot \rceil$ represents the round-up operator. $\otimes$ and $\circ$  represent the Kronecker product  and Khatri-Rao (KR)
product, respectively.    $\text{vec}(\cdot)$ represents vectorization. $\mathbf{I}$ and $\mathbf{0}$ denote an identity matrix and an all-zero vector with appropriate dimensions, respectively. $\mathbb{C}^{n\times m}$ denotes the space of $n\times m$ complex matrices. $\mathbf 1_m \in \mathbb R^{m\times1}$ denotes the all-one column vector of length $m$. $\Gamma(o; a, b)$ denotes a Gamma distribution at the variable $o$ with shape parameter $a$ and rate parameter $b$. $\langle\cdot\rangle$ denotes the expectation (mean) operator with respect to the underlying randomness.
\section{System Model}
\begin{figure}[t]
	\centering
	\includegraphics[width=3.5in]{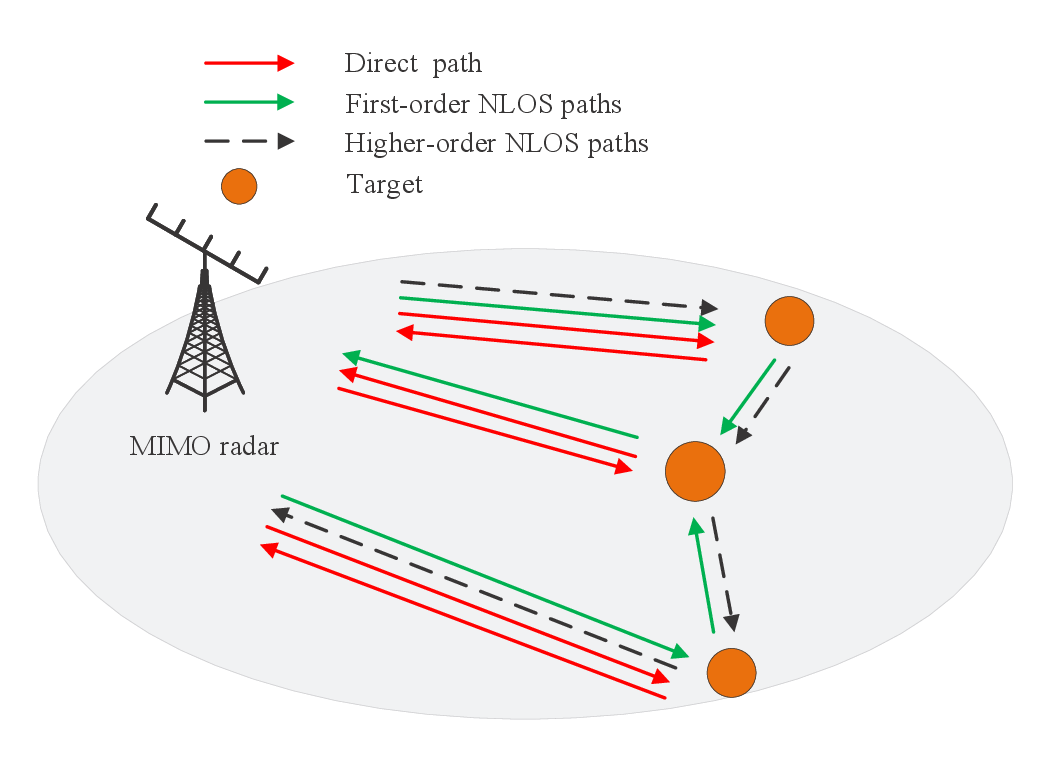}
	\caption{System model of multi-target sensing under multipath propagation.}
	\label{fig8}
\end{figure}

As shown in Fig.~\ref{fig8}, we consider a colocated MIMO radar equipped with $M_t$ transmit and $M_r$ receive antennas, both configured as uniform linear arrays (ULAs) with half-wavelength spacing, where $\lambda$ denotes the carrier wavelength. Suppose there are $K$ targets in the sensing field of view, whose direction angles  are denoted by $\boldsymbol{\theta}_T=[\theta_1,\theta_2,\cdots,\theta_K]^T$.  The radar echoes propagate among different targets and are reflected back to the receiver through multiple paths, thereby forming a multipath environment.

In such a multi-target multipath scenario, the received signal at the radar consists of both LoS and NLoS components. 
The LoS component, also referred to as the direct path, corresponds to the direct propagation between the radar and each target. In this case, the AoD and  AoA coincide with each other, as illustrated by the red line in Fig.~\ref{fig8}. Beyond these direct paths, additional multipath components arise due to the reflections among the targets. In particular, first-order NLoS paths are generated when the transmitted signal is reflected by one target before reaching another, resulting in departure and arrival angles that are associated with different targets, as illustrated by the green lines. These inter-target reflections create additional structured paths that overlap with the direct-path angles of different targets, which, if properly exploited, can provide valuable prior information for target estimation. By contrast, higher-order NLoS paths involve two or more reflections, which undergo significant attenuation and are therefore assumed negligible in this work\footnote{The residual low-energy multipath components and the radar clutters can be modeled as colored interference with an unknown covariance matrix. Within the proposed VBI framework, it is in principle possible to explicitly incorporate this colored interference and automatically learn its covariance matrix from data, thereby adaptively capturing and suppressing the clutter/multipath energy, which are left for future work.}, as indicated by the dashed black  lines.

%Since a first-order path undergoes only a single reflection, its power is generally non-negligible in sensing problems. Moreover, it carries both the departure and arrival angle information associated with two different targets. Therefore, in multi-target sensing, the exploitation of first-order paths can effectively enhance the overall sensing performance.
Target sensing aims at detecting the presence of the targets
and estimating the target locations. Let $\mathbf{u}(l)=[u_1(l),u_2(l),\cdots,u_{M_t}(l)]^T$ denote the transmitted signal at the $l$-th epoch \cite{zheng_detection_2024}. Then, the transmitted signal matrix can be represented  as $\mathbf{U}=[\mathbf{u}(1),\mathbf{u}(2),\cdots,\mathbf{u}(L)]\in\mathbb{C}^{M_t\times L}$, and the corresponding  received signal matrix  can be expressed as $\mathbf{R}=[\mathbf{r}(1),\mathbf{r}(2),\cdots,\mathbf{r}(L)]\in\mathbb{C}^{M_r\times L}$, where
\begin{equation}\label{eq:signal_model}
	\begin{aligned}
		\mathbf{r}(l)=&\underbrace{\sum_{k=1}^{K} \alpha_k \mathbf{a}_r(\theta_k)\mathbf{a}_t^{T}(\theta_k) \mathbf{u}(l)}_{\text{Direct path}}+\!\!\!\!\underbrace{\sum_{\substack{i=1,\, j=1 \\ i \neq j}}^{K} 
			\!\!\!\beta_{i,j}\mathbf{a}_r(\theta_i)\mathbf{a}_t^{T}(\theta_j) \mathbf{u}(l)}_{\text{First-order paths}}\\&+\mathbf{n}_r(l),
	\end{aligned}
	\vspace{-1mm}
\end{equation}
where $\alpha_k = \zeta_k g_k$ denotes the complex gain of the direct LoS path, with $\zeta_k \sim \mathcal{CN}(0,1)$ and $g_k$ modeling the small-scale fading and large-scale attenuation, respectively. The coefficient $\beta_{i,j} = \epsilon_{i,j} h_{i,j}$ represents the complex gain of the first-order NLoS path associated with the ordered pair $(i,j)$, where $i$ denotes the index of the AoD corresponding to the $i$-th target and $j$ denotes the index of the AoA corresponding to the $j$-th target. Here, $\epsilon_{i,j} \sim \mathcal{CN}(0,1)$ accounts for the small-scale fading, and $h_{i,j}$ represents the corresponding large-scale attenuation.
The term $\mathbf{n}_r(l)$ denotes the additive white Gaussian noise (AWGN).   $\mathbf{a}_t(\theta)=[1,e^{j {\pi}  \sin(\theta)},\cdots,e^{j {\pi}(M_t-1) \sin(\theta)}]^T\in\mathbb{C}^{M_t\times1}$ and $\mathbf{a}_r(\phi)=[1,e^{j {\pi} \sin(\phi)},\cdots,e^{j {\pi}(M_r-1)\sin(\phi)}]^T \in\mathbb{C}^{M_r\times1}$ are defined as the transmit and receive steering vectors, respectively, where $\theta$ and $\phi$ denote the transmit and receive direct angles, respectively.

\section{Sparse Bayesian Inference Formulation}
In this section, we first construct a sparse representation model of the sensing channels for the considered multi-target scenario with multipath components, based on 2D angle grids. Next, we introduce a hierarchical prior model to characterize the sensing channel sparsity, which facilitates the subsequent  radar sensing algorithm design. In particular, motivated by the characteristics of the first-order paths, we propose a cross sparsity structure based on the Ising model, which explicitly exploits the relation between the first-order  and  direct paths to resolve angular ambiguities in the estimation process, thereby effectively enhancing the estimation performance.
\subsection{Sparse Representation of the Multi-Target Sensing Channel}
To obtain a high-resolution sparse representation of the sensing channel, we uniformly discretize
the transmit and receive angle domains  over
$[-\frac{\pi}{2},\,\frac{\pi}{2}]$ into $Q$ grid points, denoted by
$\bm{\bar{\theta}}_t=[\bar{\theta}_{t,1},\cdots,\bar{\theta}_{t,Q}]^T$ and
$\bm{\bar{\theta}}_r=[\bar{\theta}_{r,1},\cdots,\bar{\theta}_{r,Q}]^{ T}$, respectively. Besides, the first-order and direct path  parameters are updated independently to improve the estimation accuracy. Accordingly, the transmit and receive angle grids are expanded to $Q_1\triangleq Q^2$ points, enabling separate refinement of the two types of paths in the 2D angular domain, i.e., $\bm{{{\theta}}}_t=\mathbf 1_Q\otimes\bm{\bar{\theta}}_t\in\mathbb{R}^{Q_1\times1}$ and $\bm{{{\theta}}}_r=\bm{\bar{\theta}}_r\otimes\mathbf 1_Q\in\mathbb{R}^{Q_1\times1}$. Let $q_{c,(i-1)K+j}\triangleq\arg\min_{q}\big(|\theta_i-{\theta}_{t,q}|+|\theta_j-{\theta}_{r,q}|\big)$, where $i,j\in\mathcal{K}\triangleq\{1,2,\cdots,K\}$ denote the index of the predefined transmit–receive angle grid pair that is nearest to the transmit angle $\theta_i$ and the receive angle $\theta_j$. In particular, the case $i=j$ corresponds to the direct path, while $i\neq j$ represents the first-order paths. We assume that $Q_1$
is sufficiently large so that different paths have distinct nearest grid points. Accordingly, for each transmit grid point $\theta_{t,q}$ and receive grid point $\theta_{q}$, $\forall q \in \mathcal{Q}^2 \triangleq \{1,2,\cdots,Q_1\}$, we define the corresponding angular offset as the difference between the actual target angle and its nearest predefined grid point,
i.e.
\begin{equation}
	{\Delta \theta}_{t,q}\triangleq
	\begin{cases}
		\theta_{\lceil \frac{q}{K}\rceil} - \theta_{t,q}, & \text{if } q \in\mathcal{Q}_c\triangleq\{q_{c,1},q_{c,2},\cdots,q_{c,K^2}\},\\[6pt]
		0, & \text{if } q \notin \mathcal{Q}_c,\\
	\end{cases}
\end{equation}
and
\begin{equation}
	{\Delta\theta}_{r,q} \triangleq
	\begin{cases}
		\theta_{q-\lceil \frac{q}{K}\rceil K} - \theta_{r,q}, & \text{if } q \in\mathcal{Q}_c,\\[6pt]
		0, & \text{if } q \notin\mathcal{Q}_c.
	\end{cases}
\end{equation}
We further define
$
\bm{\Delta\theta}_t \triangleq [\Delta\theta_{t,1},\cdots,\Delta\theta_{t,Q_1}]^{ T} $ and $
\bm{\Delta\theta}_r \triangleq [\Delta\theta_{r,1},\cdots,\Delta\theta_{r,Q_1}]^{T}
$
as the corresponding  angular offset vectors, which enable  a refined representation of the target angles by representing  each true transmit/receive angle as the sum of its nominal grid angle and the corresponding offset, and play a crucial role in achieving super-resolution sensing  throughout this paper.

Based on the above definitions,  the  sparse dictionaries with respect to $\bm{{\theta}}_t$ and $\bm{{\theta}}_r$ can be expressed as 
$
\mathbf{A}_t(\bm{\Delta\theta}_t)
= \big[\mathbf{a}_t({\theta}_{t,1}+\Delta\theta_{t,1}),\, \cdots,\, \mathbf{a}_t({\theta}_{t,Q_1}+\Delta\theta_{t,Q_1})\big]\in
\mathbb{C}^{M_t\times Q_1},
$
and
$
\mathbf{A}_r(\bm{\Delta\theta}_r)
= \big[\mathbf{a}_r({\theta}_{r,1}+\Delta\theta_{r,1}),\, \cdots,\, \mathbf{a}_r({\theta}_{r,Q_1}+\Delta\theta_{r,Q_1})\big]\in
\mathbb{C}^{M_r\times Q_1}.
$ Then, the sparse representation of the received signal in \eqref{eq:signal_model} can be expressed as 
\begin{equation}
	\mathbf{R}
	= \mathbf{H} \mathbf{U}
	+ \mathbf{N}_r,
\end{equation}
 where
$
\mathbf{H} \;=\; \mathbf{A}_r(\bm{\Delta\theta}_r)\,\text{diag}(\mathbf{x})\,\mathbf{A}_t^{T}(\bm{\Delta\theta}_t)\in\mathbb{C}^{M_r\times M_t}$,  $\mathbf{x}\in\mathbb{C}^{Q_1\times 1}$ denotes the corresponding sparse sensing channel vector, and $\mathbf{N}_r=[\mathbf{n}_r(1),\mathbf{n}_r(2),\cdots,\mathbf{n}_r(L)]\in\mathbb{C}^{M_r\times L}$.
At the radar receiver, the received signal is processed using a matched filter to   estimate the target positions, and the output of this filtering process is given by  $\mathbf{Y}=\mathbf{R}\mathbf{U}^H$.
For ease of exploitation, we propose to vectorize $\mathbf{Y}$ as
\begin{equation}\label{eq:1014_1}
\mathbf{y}=\text{vec}\{\mathbf{Y}\}
=\mathbf{F}\left(\bm{\Delta\theta}_t,\bm{\Delta\theta}_r\right)
\mathbf{x}\,
+ \mathbf{n}\in\mathbb{C}^{M_tM_r\times 1},
\end{equation}
where  
\begin{equation}\label{eq:1014_2}
\mathbf{F}\!\left(\bm{\Delta\theta}_t,\bm{\Delta\theta}_r\right)
= ((\mathbf{U}\mathbf{U}^H)^T\otimes \mathbf{I}_{M_r})(\mathbf{A}_t\!\left(\bm{\Delta\theta}_t\right) \circ \mathbf{A}_r\!\left(\bm{\Delta\theta}_r\right))
\end{equation}
denotes the sensing measurement matrix, and $\mathbf{n}=\text{vec}\{\mathbf{N}_r\mathbf{U}^H\}$.

\subsection{Probability Model under the Proposed Cross Sparsity Prior}
Building upon the sparse representation model \eqref{eq:1014_1}, we now develop a probabilistic framework, which provides a principled way to model structured sparsity, enabling Bayesian inference and the incorporation of physically motivated dependencies that deterministic models could not capture. Within this framework, we propose a novel cross sparsity structure that exploits the relation between direct and first-order paths. To instantiate this idea while remaining robust to imperfect prior information, we develop a 3LHS prior that flexibly captures the cross sparsity pattern.

%Specially, let $\boldsymbol{\rho} \triangleq [\rho_{r,1},\rho_{r,2},\cdots,\rho_{r,Q^2}]^T$ represent
%the precision vectors of $\mathbf{x}$, where $1./\rho_{q}$ is the variance  of $x_{q}$. Let $ \mathbf{s} \triangleq [s_{r,1},s_{r,2} \cdots, s_{r,Q^2}]^T \in \{0,1\}^{Q^2},$ represent the support vectors of $\mathbf{x}$. Since the diagonal elements of $\mathbf{X}_r$ correspond to the sparse signals of the direct paths, while the off-diagonal elements correspond to the sparse signals of the first-order paths, and $\mathbf{x}$ is obtained by vectorizing $\mathbf{X}_r$, it follows that $x_{r,(i-1)Q+j},i\in\{1,2,\cdots,Q\},j\in\{1,2,\cdots,Q\}$ with $i \neq j$ represents a first-order path sparse signal, whereas $x_{r,(i-1)Q+j}$ with $i=j$ represents a direct path sparse signal. Therefore, if there is a direct radar path in $(i-1)Q+i$-th grid , we have $s_{r,(i-1)Q+i}=1$ and $x_{r,(i-1)Q+i}$ is non-zero. Otherwise, we have $s_{r,(i-1)Q+i}=0$ and $x_{r,(i-1)Q+i}=0$. If there is a first order path in $(i-1)Q+j,i\neq j$, we  have $s_{r,(i-1)Q+j}=1$ and $x_{r,(i-1)Q+j}$ is non-zero. Otherwise, we have $s_{r,(i-1)Q+j}=0$ and $x_{r,(i-1)Q+j}=0$. 

Specifically, let $\boldsymbol{\rho} \triangleq [\rho_{1}, \rho_{2}, \cdots, \rho_{Q_1}]^T$ denote 
the precision vector associated with $\mathbf{x}$, where $1/\rho_{q}$ represents the variance of $x_{q}$. 
Define a   binary support vector 
$\mathbf{s} \triangleq [s_{1}, s_{2}, \ldots, s_{Q_1}]^T \in \{0,1\}^{Q_1}$ 
to indicate the activity of the entries in $\mathbf{x}$. 
The support indicator $s_{(i-1)Q+j}$ takes the value of $1$ when the corresponding coefficient 
$x_{(i-1)Q+j}$ is nonzero, and is set to zero  otherwise. 
%In particular, $s_{r,(i-1)Q+i}=1$ indicates the 
%existence of a direct path in the $(i-1)Q+i$-th grid, while $s_{r,(i-1)Q+j}=1$ with $i \neq j$ 
%indicates the presence of a first-order path. 
Then we can obtain the joint distribution of $\mathbf{x},\boldsymbol{\rho}$ and $\mathbf{s}$ as
\begin{equation}\label{eq:1030_1}
	p(\mathbf{x},\boldsymbol{\rho},\mathbf{s})=p(\mathbf{x}|\boldsymbol{\rho})p(\boldsymbol{\rho}|\mathbf{s})p(\mathbf{s}).
\end{equation}

Based on \eqref{eq:1030_1}, we can see that the sparse signal $\mathbf{x}$ follows  complex Gaussian 
distribution with zero mean and covariance matrix $\bm{\rho}^{-1}$, which preserves tractability within the CS framework 
while adaptively inducing sparsity through the hyperparameters $\bm{\rho}$ and $\mathbf{s}$.
Conditioned on $\boldsymbol{\rho}$, the prior distribution of $\mathbf{x}$ is
\begin{equation}
	p(\mathbf{x} | \boldsymbol{\rho}) 
	= \prod_{q=1}^{Q_1} p(x_{q} | \rho_{q}) 
	= \prod_{q=1}^{Q_1} \mathcal{CN}(x_{q}; 0, 1/\rho_{q}).
\end{equation}
Since  $\mathbf{x}$ is modeled as conditionally complex Gaussian distributed given the element-wise precisions $\bm{\rho}$, we propose to assign $\bm{\rho}$ a Gamma prior such that a conjugate model can be obtained that enables closed-form updates and effective sparsity-promoting shrinkage, i.e.,
\begin{equation}
p(\boldsymbol{\rho} | \mathbf{s}) 
= \prod_{q=1}^{Q_1} p(\rho_{q} | s_{q}),
\label{eq:26}
\end{equation}
where 
\begin{equation}
	\begin{aligned}
	&p(\rho_{q} | s_{q}) 
	= 
	\Big( \Gamma(\rho_{q}; a_{q}, b_{q}) \Big)^{s_{q}}
	\Big( \Gamma(\rho_{q}; \bar{a}_{q}, \bar{b}_{q}) \Big)^{1-s_{q}},
	\end{aligned}
\end{equation}
To control the amplitude of $x_{q}$ according to the support indicator $s_{q}$, 
the parameters $a_{q}$ and $b_{q}$ are chosen such that
$
\frac{a_{q}}{b_{q}} = \langle\rho_{q}\rangle 
= \mathcal{O}\!\left(1\right).
$
When $s_{q}=0$, $x_{q}$ is equal to zero or close to zero,
in this case, the corresponding parameters $\bar{a}_{q}$ and $\bar{b}_{q}$ are set such that 
$
\frac{\bar{a}_{q}}{\bar{b}_{q}} = \langle\rho_{q}\rangle \gg 1,
$
ensuring that the variance of $x_{q}$ approaches zero.
In addition, the  noise precision $\gamma$,  defined as the inverse of the noise variance, is also modeled as a Gamma distributed variable  with parameters $c$ and $d$, i.e.,
\begin{equation}
p(\gamma) = \mathrm{\Gamma}(\gamma; c, d).
\label{eq:noise_gamma}
\end{equation}

\begin{figure}[t]
	\centering
	\includegraphics[width=3.5in]{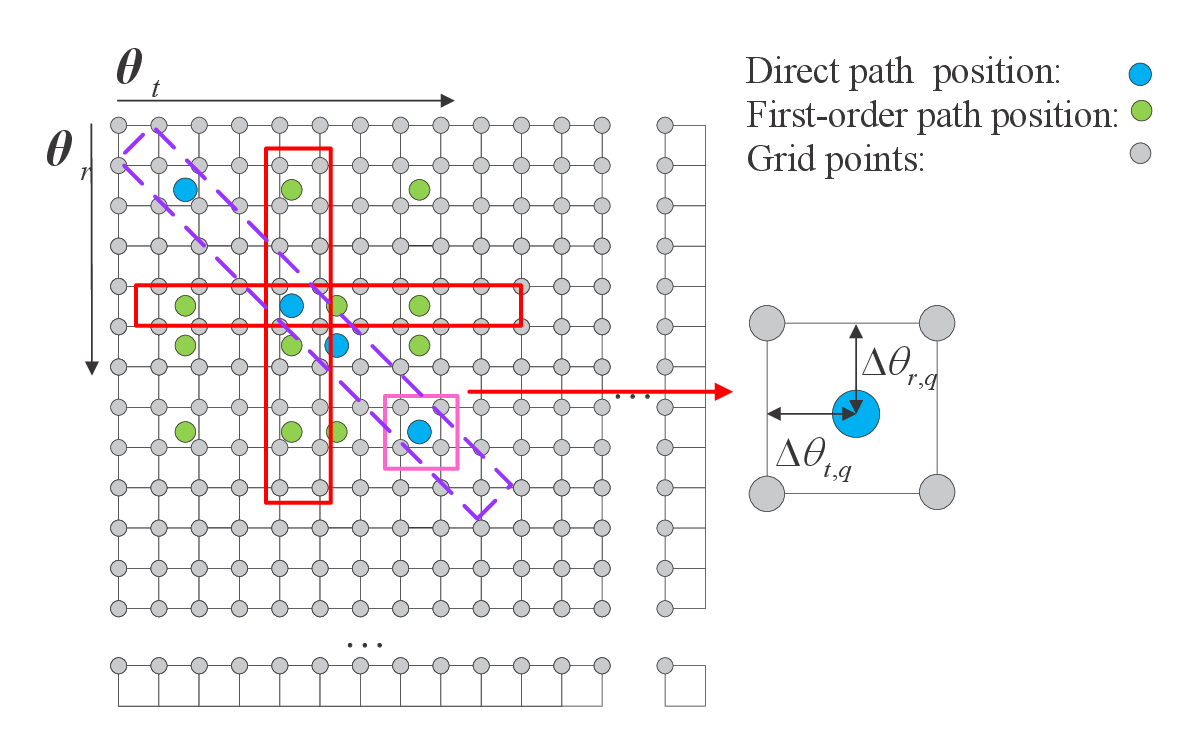}
	\caption{Illustration of cross sparsity.}
	\label{fig9}
\end{figure}

In \eqref{eq:1030_1}, the prior distribution $p(\mathbf{s})$ is used to capture the underlying  structured sparsity and  the dependency within the support vector $\mathbf{s}$. While Bernoulli and Markov priors are effective for independent and clustered structures \cite{liu_robust_2020,11024132}, respectively, they cannot characterize the special property of  multipath scenarios with distinct transmit and receive angles or exploit first-order paths to enhance direct path perception. To address this limitation, we propose a novel {cross sparsity} structure that explicitly leverages the existence of first-order paths. Specifically,  since the AoD and AoA of a first-order path coincide with the direct path angles of different targets, the presence of such a path increases the prior probability that the corresponding direct paths are active at those transmit/receive angles, as illustrated in Fig. \ref{fig9}. The direct paths exhibit equal transmit and receive angles, and hence their corresponding sparse coefficients appear along the main diagonal. In contrast, the first-order paths share the transmit or receive angle with the direct paths of  different targets, leading to off-diagonal sparse coefficients. Consequently, the overall sparsity  forms a cross-shaped structure, which we refer to as {cross sparsity}.

To effectively capture the cross sparsity structure of the considered multi-target sensing channel, 
we model the prior distribution of the  binary support vector $\mathbf{s}$ using an Ising model \cite{xu_joint_2024}, which can be expressed as
\begin{equation}\label{eq:0822_2}
	\begin{aligned}
		p(\mathbf{s};\boldsymbol{\omega})
		= \frac{1}{z(\boldsymbol{\omega})}
		\prod_{i\neq j}
		&\varphi\big(s_{i(Q-1)+j}, s_{i(Q-1)+i}\big)\\&
		\varphi\big(s_{i(Q-1)+j}, s_{j(Q-1)+j}\big),
	\end{aligned}
\end{equation}
where the partition function
\begin{equation}
	\begin{aligned}
		z(\boldsymbol{\omega})
		\triangleq \sum_{\mathbf{s}}
		\prod_{i\neq j}
		&\varphi\big(s_{i(Q-1)+j}, s_{i(Q-1)+i}\big)\\&
		\varphi\big(s_{i(Q-1)+j}, s_{j(Q-1)+j}\big),
	\end{aligned}
\end{equation} 
ensures that $\sum_{\mathbf{s}} p(\mathbf{s};\boldsymbol{\omega}) = 1$, 
and
\begin{equation}
	\begin{aligned}
		&\varphi\big(s_{i(Q-1)+j}, s_{i(Q-1)+i}\big) \\
		&\triangleq 
		e^{\omega_{i(Q-1)+j,i(Q-1)+i}\, s_{i(Q-1)+j}, s_{i(Q-1)+i}}, \quad 
		\forall i,j\in\mathcal{Q},\, i\neq j,
	\end{aligned}
\end{equation}
denotes the potential function defined over the first-order path support variable $s_{i(Q-1)+j}$ and 
the direct-path support variable $s_{i(Q-1)+i}$. Here, 
$
\boldsymbol{\omega}
\triangleq
\big\{
\omega_{i(Q-1)+j,\,i(Q-1)+i},
\,
\omega_{i(Q-1)+j,\,j(Q-1)+j}
\mid
i,j\in\mathcal{Q},\, i\neq j
\big\}
$
collects the interaction parameters that characterize the correlation strength associated with the
cross sparsity. In practice, these interaction parameters are not manually specified; instead,
$\boldsymbol{\omega}$ is treated as a set of learnable hyperparameters and is automatically
updated during the M-step of the proposed SF-VBI algorithm.

Based on the above hierarchical prior model, the joint
distribution of all the considered random variables can be
written as
\begin{equation}\label{eq:0822_4}
	\begin{aligned}
		&p(\mathbf{y},\mathbf{x},\boldsymbol{\rho},\mathbf{s},\gamma;\boldsymbol{\xi})\\&=p(\mathbf{y}|\mathbf{x},\gamma;\boldsymbol{\Delta{\theta}}_t,\boldsymbol{\Delta{\theta}}_r)p(\mathbf{x}|\boldsymbol{\rho})p(\boldsymbol{\rho}|\mathbf{s})p(\mathbf{s};\boldsymbol{\omega})p(\gamma),
	\end{aligned}
\end{equation}
where $\boldsymbol{\xi}$ represents the set of parameters expressed as  $\boldsymbol{\xi}\triangleq\{\boldsymbol{\Delta\theta}_t,\boldsymbol{\Delta\theta}_r, \boldsymbol{\omega}\} $, and $ p(\mathbf{y}|\mathbf{x},\gamma; \boldsymbol{\Delta{\theta}}_t,\boldsymbol{\Delta{\theta}}_r) 
= \mathcal{CN}\!\left({\mathbf{F}}(\boldsymbol{\Delta{\theta}}_t,\boldsymbol{\Delta{\theta}}_r)\mathbf{x}, \gamma^{-1}\mathbf{I}\right)$.
With given observation vector $\mathbf{y}$ and parameter set 
$\boldsymbol{\xi}$, our main objective is to calculate 
the accurate marginal posteriors $p(\mathbf{x}|\mathbf{y}; \boldsymbol{\xi})$, 
$p(\boldsymbol{\rho}|\mathbf{y}; \boldsymbol{\xi})$, 
$p(\mathbf{s}|\mathbf{y}; \boldsymbol{\xi})$ and $p(\gamma|\mathbf{y}; \boldsymbol{\xi})$ by performing 
Bayesian inference for $\mathbf{x}$, $\boldsymbol{\rho}$, $\mathbf{s}$ and $\gamma$, respectively. 
Furthermore, the optimal parameter set $\boldsymbol{\xi}^*$ 
can be determined by solving the following maximum likelihood (ML) problem:
\begin{equation}\label{eq:0825_1}
\boldsymbol{\xi}^*
	= \arg\max_{ \boldsymbol{\xi}} 
	\ln p(\mathbf{y}| \boldsymbol{\xi}).
\end{equation}
However, the likelihood function $\ln p(\mathbf{y}| \boldsymbol{\xi})$  does not admit a closed-form expression as it requires multidimensional integration over the latent variables $\mathbf{x}, \bm{\rho}, \mathbf{s}$ and $\gamma$, and the loops in the factor graph of model \eqref{eq:0822_4} render exact inference intractable, making it challenging to obtain accurate marginals for $\mathbf{x}, \bm{\rho}, \mathbf{s}$ and $\gamma$. To address these issues, the Turbo-VBI algorithm~\cite{liu_robust_2020} that combines VBI, message passing, and EM  can be employed  to approximate the marginal posteriors and ML estimates, but its alternating E–M steps are too costly for real-time sensing. Motivated by the two-timescale alternating MAP method in~\cite{zhou_robust_2025}, we propose in the following an SF-TVBI algorithm that significantly lowers the computational complexity while preserving estimation accuracy.

\section{SF-TVBI Algorithm}
\begin{figure}[t]
	\centering
	\includegraphics[width=3.5in]{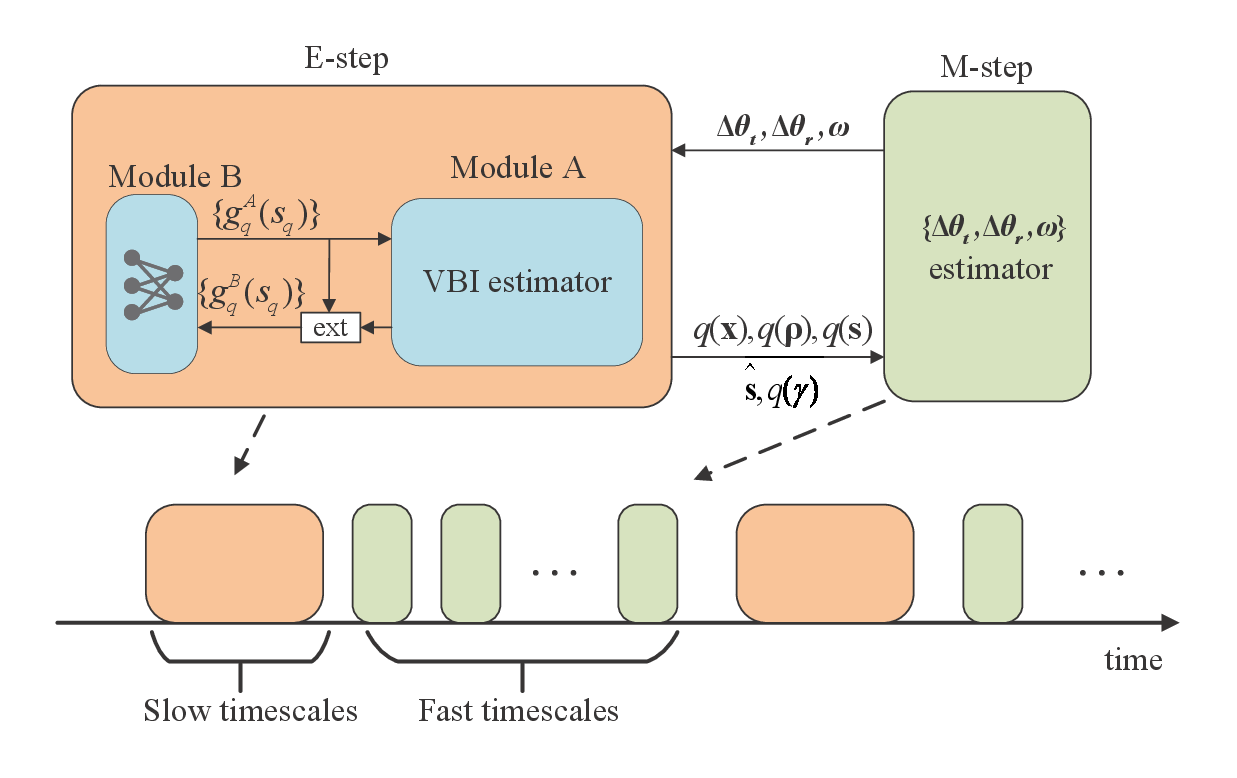}
	\caption{Overall framework of the proposed SF-TVBI algorithm.}
	\label{fig5}
\end{figure}

In this section, we present the SF-TVBI algorithm to address the above estimation problem. The algorithm is developed under the EM framework~\cite{liu_robust_2020}, consisting of an E step and an M step, and adopts a two-timescale strategy, where the M step runs in inner loop at a faster time-scale, while the E step proceeds in a  outer loop at a slower time-scale. This design effectively reduces the number of E step iterations, thereby achieving lower complexity than that of Turbo-VBI \cite{liu_robust_2020}.  
In the E step, given $\mathbf{y}$ and the parameter set $\boldsymbol{\xi}$, the marginal posteriors of $\mathbf{x}$ and $\boldsymbol{\rho}$ are approximated by tractable distributions via VBI. To cope with the excessive loops in the factor graph induced by the cross sparsity structure, an efficient message passing algorithm is employed to compute the posterior probability of the support vector $\mathbf{s}$, denoted by $q(\mathbf{s})$. This probability acts as the discriminant for determining the estimated support vector $\hat{s}$ via a hard-thresholding rule with threshold $\tau_s$, i.e.,
\begin{equation}\label{eq:0904_1}
	\hat{s}_{q} =
	\begin{cases}
		1, & q(s_{q}) > \tau_s, \\
		0, & \text{otherwise},
	\end{cases}
	\qquad q \in \mathcal{Q}^2\triangleq\{1,2,\cdots,Q_1\},
\end{equation}  
thereby identifying the set of grid points for subsequent update in the M-step. In the M step, a surrogate function  of $\ln p(\mathbf{y}|\boldsymbol{\xi})$ is constructed using the E step outputs, and a gradient-ascent (GA) solver is applied when updating the selected grids. For clarity, the overall flow of the proposed algorithm is summarized in Fig.~\ref{fig5}. Note that compared  with the two-timescale alternating MAP method~\cite{zhou_robust_2025}, which relies on the sparse signal energy  as the update criterion, the proposed probability-based strategy is less sensitive to the signal and noise power, and thus  provides enhanced robustness.  

\subsection{SF-TVBI-E Step}
Due to the numerous loops in the factor graph of the joint 
distribution~\eqref{eq:0822_4}, computing the exact marginal posteriors with 
respect to $\mathbf{x}$, $\boldsymbol{\rho}$, $\mathbf{s}$ and $\gamma$ is an 
NP-hard problem \cite{chen_joint_2024}. To address this issue, we approximate the marginal posteriors 
by partitioning the original factor graph into two sub-graphs corresponding to 
Module~A and Module~B, as illustrated in Fig.~\ref{fig5}. Module A is
responsible for approximating the marginal posteriors of $\mathbf{x}$,
$\bm{\rho}$,  $\mathbf{s}$ and $\gamma$ using the VBI method, while Module B focuses
on updating the support prior by processing the cross sparsity
structure. The two modules exchange messages in an iterative
manner to enable tractable inference. In particular, 
given the observation vector $\mathbf{y}$ and the output messages from 
Module~B, denoted by $\{g^{A}_{q}(s_{q})\}$, $\forall q \in 
\mathcal{Q}^2$ (i.e., extrinsic prior probabilities from the cross sparsity), Module~A employs the VBI method to compute the approximate 
marginal posteriors $q(\mathbf{x})$, $q(\boldsymbol{\rho})$, and 
$q(\mathbf{s})$ under the following prior distribution:
\begin{equation}
	\hat{p}(\mathbf{x}, \boldsymbol{\rho}, \mathbf{s},\gamma) 
	= p(\mathbf{x} | \boldsymbol{\rho})\,
	p(\boldsymbol{\rho} | \mathbf{s})\,
	\hat{p}(\mathbf{s})p(\gamma),
	\label{eq:0825_2}
\end{equation}
where $\hat{p}(\mathbf{s})=\prod_{q=1}^{Q^2}(\pi_{q})^{s_{q}}
(1-\pi_{q})^{1-s_{q}}$ with
\begin{equation}
	\pi_{q}=\frac{g^{A}_{q}(1)}{g^{A}_{q}(1)+g^{A}_{q}(0)}.
	\label{eq:0825_3}
\end{equation}
The output messages from Module~A (i.e., extrinsic pseudo-likelihoods serving as next-iteration priors), 
denoted by $\{g^{B}_{q}(s_{q})\}$, $\forall q \in \{1,2,\ldots,Q_1\}$, can 
be calculated as
\begin{equation}
	g^{B}_{q}(s_{q}) = \frac{q(s_{q})}{g^{A}_{q}(s_{q})} \propto \pi^{B}_{q}\delta(s_{q}-1)+(1-\pi^{B}_{q})\delta(s_{q}),
	\label{eq:34}
\end{equation}
where  $\pi^{B}_{q}$ is the posterior existence probability for the $q$-th grid. Using $\{g^{B}_{q}(s_{q})\}$ as priors, Module B executes the sum-product message passing algorithm on the factor graph defined by the cross sparsity structure to generate the updated messages $\{g^{A}_{q}(s_{q})\}$, Modules A and B then exchange these messages iteratively until a stable convergence criterion is met.
The details of these two modules are presented as follows.

1) \textit{VBI Estimator (Module A)}:
The goal of VBI estimation  is to compute the approximate marginal posteriors 
$q(\mathbf{w}), \, \forall \mathbf{w} \in \Omega \triangleq 
\{\mathbf{x}, \boldsymbol{\rho}, \mathbf{s},\gamma\}$ 
by minimizing the Kullback--Leibler divergence (KLD) between 
$q(\mathbf{x}, \boldsymbol{\rho}, \mathbf{s},\gamma) 
= \prod_{\mathbf{w} \in \Omega}q(\mathbf{w})$ 
and the posterior distribution 
$\hat{p}(\mathbf{x}, \boldsymbol{\rho}, \mathbf{s},\gamma |\mathbf{y}; 
	\boldsymbol{\Delta{\theta}}_t,\boldsymbol{\Delta{\theta}}_r)$
under the prior~\eqref{eq:0825_2}, i.e.,
\begin{equation}
	\min_{q(\mathbf{x},\boldsymbol{\rho},\mathbf{s},\gamma)} \text{KL}(q(\mathbf{x},\boldsymbol{\rho},\mathbf{s},\gamma)||\hat{p}(\mathbf{x},\boldsymbol{\rho},\mathbf{s},\gamma|\mathbf{y};	\boldsymbol{\Delta{\theta}}_t,\boldsymbol{\Delta{\theta}}_r),
	\label{eq:36}
\end{equation}
where
\begin{equation}\label{eq:0825_4}
	\begin{aligned}
	&\text{KL}(q(\mathbf{x},\boldsymbol{\rho},\mathbf{s},\gamma)||\hat{p}(\mathbf{x},\boldsymbol{\rho},\mathbf{s},\gamma|\mathbf{y};	\boldsymbol{\Delta{\theta}}_t,\boldsymbol{\Delta{\theta}}_r))\\&\triangleq	\int 
		\ln \frac{q(\mathbf{x},\boldsymbol{\rho},\mathbf{s},\gamma)}
		{\hat{p}(\mathbf{x},\boldsymbol{\rho},\mathbf{s},\gamma|\mathbf{y};
	\boldsymbol{\Delta{\theta}}_t,\boldsymbol{\Delta{\theta}}_r)} 
			q(\mathbf{x},\boldsymbol{\rho},\mathbf{s},\gamma) 
		d\mathbf{x}\, d\boldsymbol{\rho}\, d\mathbf{s}\,d\gamma.
	\end{aligned}
\end{equation}
Although  problem \eqref{eq:36} is known to be non-convex, it is convex w.r.t. a 
single variational distribution, e.g., $q(\mathbf{w})$, after fixing the others, e.g., $q(\mathbf{v}), \, \forall \mathbf{v}\in\Omega\backslash  \mathbf{w}$, and it has been proved 
in \cite{5585756} that a stationary solution can be obtained  via optimizing each 
variational distribution in an alternating fashion. Specifically, for given 
$q(\mathbf{v}), \, \forall \mathbf{v}\in \Omega\backslash \mathbf{w}$, the optimal $q(\mathbf{w})$ that minimizes the 
KL-divergence is given by \cite{11024132}
\begin{equation}
	\begin{aligned}
		&q(\mathbf{w}) \propto \exp\!\left(
		\left\langle \ln \hat{p}(\mathbf{x}, \boldsymbol{\rho}, \mathbf{s},\gamma, \mathbf{y}; 
	\boldsymbol{\Delta{\theta}}_t,\boldsymbol{\Delta{\theta}}_r) \right\rangle
		_{\prod_{\mathbf{v}\in\Omega \setminus \mathbf{w}} q(\mathbf{v})}
		\right),
		\label{eq:37}
	\end{aligned}
\end{equation}
where $\langle \cdot \rangle_{\prod_{\mathbf{v}\in\Omega \setminus \mathbf{w}} q(\mathbf{v})}$ is an expectation 
operation w.r.t. $q(\mathbf{v})$ for $\mathbf{v}\in\Omega\backslash\mathbf{w}$.  The joint distribution $\hat{p}(\mathbf{x}, \boldsymbol{\rho}, \mathbf{s},\gamma, \mathbf{y}; 
	\boldsymbol{\Delta{\theta}}_t,\boldsymbol{\Delta{\theta}}_r)$ with  given prior distribution  $\hat{p}(\mathbf{x}, \boldsymbol{\rho}, \mathbf{s},\gamma)$ can be obtained by 
\begin{equation}\label{eq:0825_5}
	\begin{aligned}
		&\hat{p}(\mathbf{x}, \boldsymbol{\rho}, \mathbf{s}, \gamma,\mathbf{y}; 
	\boldsymbol{\Delta{\theta}}_t,\boldsymbol{\Delta{\theta}}_r)\\&=p(\mathbf{y}|\mathbf{x},\gamma;
	\boldsymbol{\Delta{\theta}}_t,\boldsymbol{\Delta{\theta}}_r)\hat{p}(\mathbf{x}, \boldsymbol{\rho}, \mathbf{s})p(\gamma),
	\end{aligned}
\end{equation}
where $p(\mathbf{y}|\mathbf{x},\gamma;
\boldsymbol{\Delta \theta}_t, \boldsymbol{\Delta \theta}_r)$ is the likelihood function, 
$\hat{p}(\mathbf{x}, \boldsymbol{\rho}, \mathbf{s})$ and $p(\gamma)$ are the priors given in \eqref{eq:0825_2}
and \eqref{eq:noise_gamma}, respectively.

By substituting \eqref{eq:0825_5} into \eqref{eq:37}, we can then obtain the variational distribution $q(\mathbf{w}),\forall \mathbf{w}\in\bm{\Omega}$. Specifically, by treating
$\boldsymbol{\Delta \theta}_t$ and $\boldsymbol{\Delta \theta}_r$ as deterministic parameters and applying
\eqref{eq:37} while discarding the terms irrelevant to $\mathbf{x}$,
the posterior distribution $q(\mathbf{x})$ is obtained as 
\begin{equation}
	\begin{aligned}
	q(\mathbf{x}) 
		=\mathcal{CN}(\mathbf{x};\boldsymbol{\mu}_\mathbf{x},\boldsymbol{\sigma}_\mathbf{x} ),
	\end{aligned}
	\label{eq:k14}
\end{equation}
where
\begin{equation}
	\boldsymbol{\mu}_\mathbf{x} 
	= \boldsymbol{\sigma}_\mathbf{x} \, \langle \gamma \rangle 
	{\mathbf{F}}(\boldsymbol{\Delta{\theta}}_t,\boldsymbol{\Delta{\theta}}_r)^H \mathbf{y},
\end{equation}
and
\begin{equation}\label{eq:15}
	\boldsymbol{\sigma}_\mathbf{x} 
	= \Big( \langle \gamma \rangle 	{\mathbf{F}}(\boldsymbol{\Delta{\theta}}_t,\boldsymbol{\Delta{\theta}}_r)^H 
	{\mathbf{F}}(\boldsymbol{\Delta{\theta}}_t,\boldsymbol{\Delta{\theta}}_r)
	+ \mathrm{diag}(\langle \boldsymbol{\rho} \rangle) \Big)^{-1}.
\end{equation}
Next, the distribution of $\boldsymbol{\rho}$ can be expressed as a
product of Gamma distributions, i.e.,
\begin{equation}\label{eq:1003_1}
	\begin{aligned}
		q(\boldsymbol{\rho}) = \prod_{n=1}^{Q_1} \mathrm{Ga}(\rho_{n};\,\tilde{a}_{n}, \tilde{b}_{n}),
	\end{aligned}
\end{equation}
with the shape and rate parameters updated as
$
\tilde{a}_{n} = \langle s_{n} \rangle a_{n} + \langle 1-s_{n} \rangle \bar{a}_{n} + 1$ and $
\tilde{b}_{n} = \langle s_{n} \rangle b_{n} + \langle 1-s_{n} \rangle \bar{a}_{n} + \langle x_{n}^2 \rangle $, respectively.
Furthermore, the distribution of $\mathbf{s}$ is given as a product of
Bernoulli distributions as shown below
\begin{equation}\label{eq:k0826_3}
	q(\mathbf{s}) = \prod_{n=1}^{Q_1} \tilde{\lambda}_n^{\,s_{n}} \left(1-\tilde{\lambda}_n\right)^{\,1-s_{n}},
\end{equation}
where
\begin{equation}\label{eq:k0826_4}
	\tilde{\lambda}_n = \frac{\pi_{n} C_{n}}{\pi_{n} C_{n} + (1-\pi_{n})\bar{C}_{n}},
\end{equation}
and the terms $C_{n}$ and $\bar{C}_{n}$ are given by
\begin{equation}
C_{n} = \frac{b_n^{a_n}}{\Gamma(a_n)}
\exp\!\left( (a_n - 1)\langle \ln \rho_{n} \rangle - b_n \langle \rho_{n} \rangle \right)
\end{equation}
and
\begin{equation}
\bar{C}_{n} = \frac{\bar{b}_n^{\bar{a}_n}}{\Gamma(\bar{a}_n)}
	\exp\!\left( (\bar{a}_n - 1)\langle \ln \rho_{n} \rangle - \bar{b}_n \langle \rho_{n} \rangle \right),
\end{equation}
respectively.
Finally, we can see that  the posterior distribution of the precision parameter $\gamma$
also follows a Gamma distribution
$
	\small
	q(\gamma) = \mathrm{Ga}(\gamma; \tilde{c}, \tilde{d}),
$
with updated parameters
\begin{equation}\label{eq:1031_1}
\tilde{c} = c + Q_1
\end{equation}
 and 
\begin{equation}\label{eq:1031_2}
	\tilde{d} = d + \left\langle \|\mathbf{y} - 	{\mathbf{F}}(\boldsymbol{\Delta{\theta}}_t,\boldsymbol{\Delta{\theta}}_r) \mathbf{x}\|^2 \right\rangle_{q(\mathbf{x})},
\end{equation}
respectively. The detailed derivations of the posterior distributions $q(\mathbf{x}),\, q(\bm{\rho}),\, q(\mathbf{s})$ and $q(\gamma)$  are provided in Appendix \ref{B}.

\begin{figure}[t]
	\centering
	\includegraphics[width=3in]{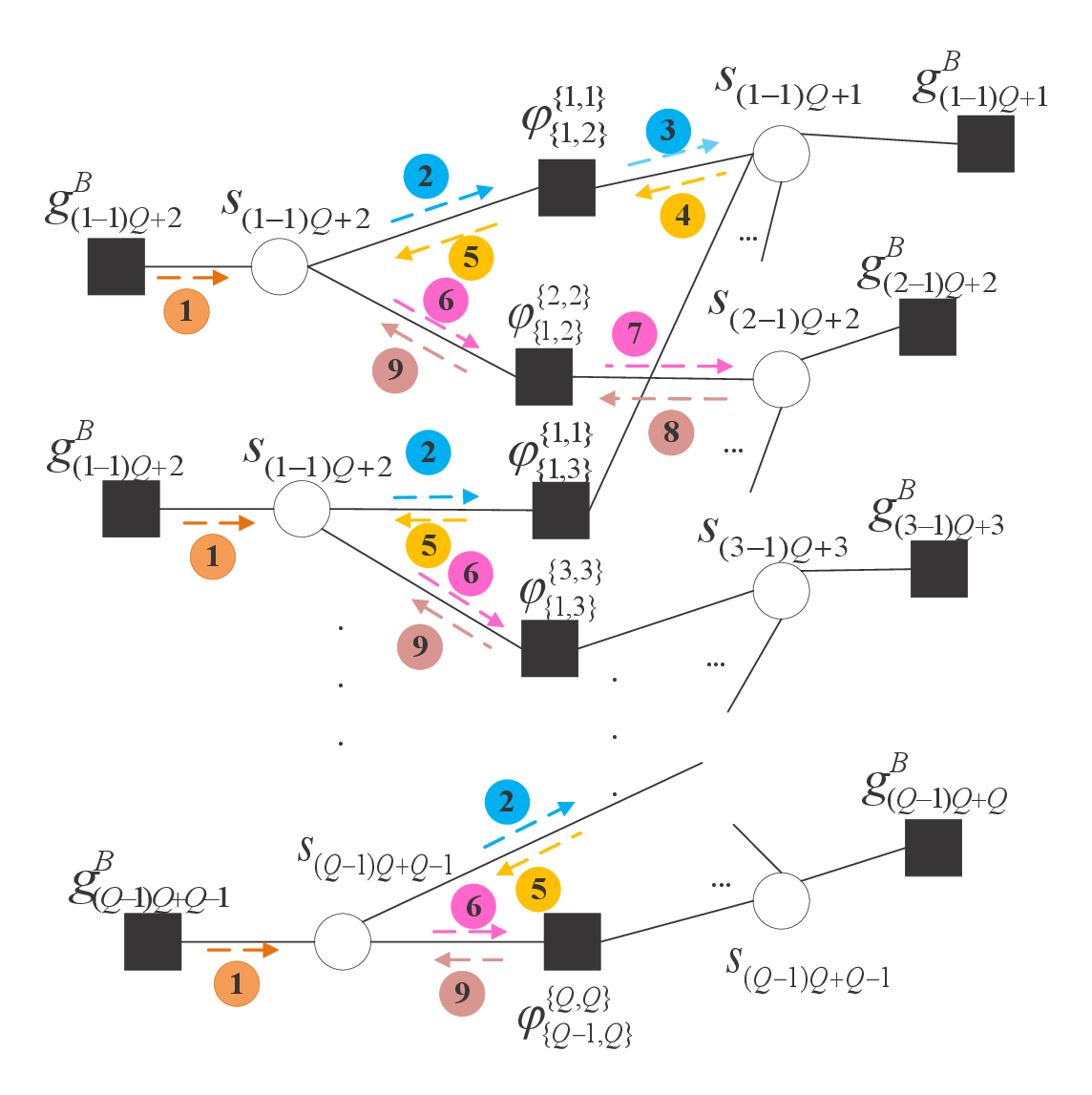}
	\caption{Factor graph of $p(\mathbf{s})$ in module B.}
	\label{fig6}
	\vspace{-3mm}
\end{figure}

2) \textit{Message Passing for Cross Sparsity (Module B)}:
The factor graph of  \( p(\mathbf{s};\boldsymbol{\omega}) \) in Module~B is illustrated in Fig.~\ref{fig6}, and the associated factor nodes along with their functional
expressions are summarized in Table I. The output messages from Module~A, i.e., \( \{g^{B}_{q}(s_{q})\} \), are incorporated as prior factor nodes within this graph.
To compute the marginal posterior distribution of the support vector \( \mathbf{s} \), sum-product message passing is performed over this factor graph. However, different from existing  sparsity models, such as block  and burst sparsity,  the proposed {cross sparsity} model induces joint row-column dependencies, leading to a loopy and structured topology, which is more complex and poses significant challenges to the application of message passing algorithms.  To address these challenges and achieve more tractable inference, we propose a simplified message passing strategy.   As can be seen in Fig. \ref{fig6}, 
%For illustration, in the type $1$ path, the message conveyed along path $1$ is proportional to the product of the messages along paths $9$ and $7$, i.e, $\nu_{s_{(i-1)Q+j}\rightarrow\varphi^{\{i,i\}}_{\{i,j\}}}\propto g^B_{(i-1)Q+j}\times \nu_{\varphi^{\{j,j\}}_{\{i,j\}}\rightarrow s_{(i-1)Q+j}} \propto\pi^{p,\{i,j\}}_{tp1}\delta(s_{(i-1)+j}-1)+(1-\pi^{p,\{i,j\}}_{tp1})\delta(s_{(i-1)+j}), \forall i,j\in\mathcal{Q},i\neq j$, where $\nu_{a\rightarrow b}$ denotes the message passed from node $a$ to node $b$, and 
%$\pi^{p,\{i,j\}}_{tp1} = 
%	\frac{\pi^{n,\{i,j\}}_{tp2}\,\pi^B_{(i-1)Q+j}}
%	{\pi^{n,\{i,j\}}_{tp2}\,\pi^B_{(i-1)Q+j} + \big(1-\pi^{n,\{i,j\}}_{tp2}\big)\big(1-\pi^B_{(i-1)Q+j}\big)}$
%. $\pi^{n,\{i,j\}}_{tp2}$ and $\pi^B_{(i-1)Q+j}$ represents the probability of $s_{(i-1)Q+j}=1$ in the message passing over the path ${\varphi_{\{i,j\}}^{\{j,j\}}\rightarrow s_{(i-1)Q+j}}$ and path $g^B_{(i-1)Q+j}\rightarrow s_{(i-1)Q+j}$, respectively.  
we classify the paths into nine types according to their connectivity relationships. Message computation starts from the nodes $s_{(i-1)Q+j}, i\neq j,i,j\in\mathcal{Q}$ with simpler connectivity relationships. We first calculate the messages transmitted along path type 1 $\rightarrow$ path type 2 $\rightarrow$ path type 3 and path type 1 $\rightarrow$ path type  6 $\rightarrow$ path type 7. It can be observed that all these messages eventually converge to the variable nodes $s_{(i-1)Q+i}$ and $s_{(j-1)Q+j}$, respectively, where $i\neq j$ and $i,j\in\mathcal{Q}$. Consequently, each variable node $s_{(i-1)Q+i}$ receives the messages from $Q$ paths of type~4 and $Q$ paths of type~7. Based on these incoming messages, the outgoing messages along path type 4 and path type 8 can be derived, which are further used to update the messages on path type 5 and path type 9. Following this message passing strategy, we are able to  iteratively update the messages along all path types until convergence. 
The detailed classification and derivation process are presented in Appendix. \ref{A}.

After completing the message updates over the support factor graph in Module~B, the output messages passed to Module~A are obtained by combining the updated beliefs from the associated cross sparsity factor nodes and the incoming prior information from Module~A, which can be expressed as
\begin{equation}\label{eq:0828_2}
	\begin{aligned}
			g^{A}_{(i-1)Q+i} \propto 
		&\left( \prod_{k=1,\,k\neq i}^{Q} \nu_{\varphi_{\{i,k\}}^{\{i,i\}} \rightarrow s_{(i-1)Q+i}} \right) 
		\\&\times 
		\left( \prod_{k=1,\,k\neq i}^{Q} \nu_{\varphi_{\{k,i\}}^{\{i,i\}} \rightarrow s_{(i-1)Q+i}} \right) 
		\\&\times g^{B}_{(i-1)Q+i}, \quad i \in \{1,2,\dots,Q\},
	\end{aligned}
\end{equation}
and
\begin{equation}
	\begin{aligned}
		g^{A}_{r,(i-1)Q+j} \propto &
		\nu_{\varphi_{\{i,j\}}^{\{i,i\}} \rightarrow s_{(i-1)Q+j}} 
	\times 
		\nu_{\varphi_{\{i,j\}}^{\{j,j\}} \rightarrow s_{(i-1)Q+j}} 
		\\&\times 
		g^{B}_{r,(i-1)Q+j}, \quad i \neq j,\ i,j \in \{1,2,\dots,Q\}.
	\end{aligned}
\end{equation}
These messages serve as extrinsic information that is fed back to Module~A and subsequently used as prior input in the next Turbo iteration.
\begin{table}[t]\label{table1}
	\centering
	\caption{Factors, Distributions and Functions Forms in Fig. \ref{fig6}}
	\label{tab:table1}
	\resizebox{\columnwidth}{!}{ % 自动缩放到单栏宽度
		\begin{tabular}{c c c}
			\hline
			\textbf{Factor node} & \textbf{Distribution} & \textbf{Detailed expression} \\
			\hline
			$\varphi^{\{i',j'\}}_{\{i,j\}}$ & 
			$\varphi(s_{(i-1)Q+j},s_{(i'-1)Q+j'})$ & 
			$e^{\omega s_{(i-1)Q+j}s_{(i'-1)Q+j'} }$ \\
			%					$\eta_q$ & 
			%					$p(\rho_{r,q} | s_{r,q})$ & 
			%					$\Big( \Gamma(\rho_{r,q}; a_{r,q}, b_{r,q}) \Big)^{s_{r,q}}
			%					\Big( \Gamma(\rho_{r,q}; \bar{a}_{r,q}, \bar{b}_{r,q}) \Big)^{1-s_{r,q}}$ \\
			%					$f_q$ & $p(x_{r,q}|\rho_{r,q}) $ & $\mathcal{CN}(x_{r,q}; 0, 1/\rho_{r,q})$\\
			$g^B_{q}$ & ${q(s_{q})}/{g^{A}_{q}(s_{q})}$ &  $\pi^{B}_{q}\delta(s_{q}-1)+(1-\pi^{B}_{q})\delta(s_{q})$\\
			\hline
		\end{tabular}
	}
\end{table}

\subsection{SF-TVBI-M Step}
In order to estimate the parameter set $\boldsymbol{\xi}$, we need 
to solve the ML problem \eqref{eq:0825_1}, which is quite challenging since there 
is no explicit expression of $\ln p(\mathbf{y} | \boldsymbol{\xi})$ 
as previously discussed. To tackle this challenge, we propose 
to construct a sequence of surrogate functions for $\ln p(\mathbf{y} | \boldsymbol{\xi})$. 
In particular, by exploiting the approximate marginal posteriors $q(\mathbf{w}), \forall \mathbf{w}\in \Omega$ 
obtained in the E step, a tractable surrogate function (in the $n$-th EM iteration) is given by
\begin{equation}\label{eq:1113_1}
	\begin{aligned}
		\mathcal{L}(\boldsymbol{\xi};\boldsymbol{\xi}^n) 
		&=\!\!\!\!\int q(\mathbf{x}, \boldsymbol{\rho}, \mathbf{s}) 
		\ln \frac{{p}(\mathbf{x}, \boldsymbol{\rho}, \mathbf{s}, \mathbf{y}; \boldsymbol{\xi})}
		{q(\mathbf{x}, \boldsymbol{\rho}, \mathbf{s})} \, d\mathbf{x}\, d\boldsymbol{\rho}\, d\mathbf{s} \\
		&=  \mathcal{L}_{\boldsymbol{\theta}}(\boldsymbol{\Delta{\theta}}_t,\boldsymbol{\Delta{\theta}}_r; \boldsymbol{\Delta{\theta}}^n_t,\boldsymbol{\Delta{\theta}}^n_r)+\mathcal{L}_{\boldsymbol{\omega}}(\boldsymbol{\omega}; \boldsymbol{\omega}^n)
		+ K,
	\end{aligned}
\end{equation}
where
\begin{equation}\label{eq:43}
	\begin{aligned}
		&\mathcal{L}_{\boldsymbol{\theta}}(\boldsymbol{\Delta{\theta}}_t,\boldsymbol{\Delta{\theta}}_r; \boldsymbol{\Delta{\theta}}^n_t,\boldsymbol{\Delta{\theta}}^n_r) \\
%		&=\!\!\!\!\int q(\mathbf{x}, \boldsymbol{\rho}, \mathbf{s}) 
%		\ln \frac{\hat{p}(\mathbf{x}, \boldsymbol{\rho}, \mathbf{s}, \mathbf{y}; \boldsymbol{\Delta{\theta}}_t,\boldsymbol{\Delta{\theta}}_r)}
%		{q(\mathbf{x}, \boldsymbol{\rho}, \mathbf{s})} \, d\mathbf{x}\, d\boldsymbol{\rho}\, d\mathbf{s} \\
		&\triangleq  - \langle \gamma \rangle \cdot 
		\bigg( \left\| \mathbf{y} -	{\mathbf{F}}(\boldsymbol{\Delta{\theta}}_t,\boldsymbol{\Delta{\theta}}_r)\boldsymbol{\mu}_\mathbf{x} \right\|^2 
		+\\&\quad \mathrm{Tr}\!\left( 	{\mathbf{F}}(\boldsymbol{\Delta{\theta}}_t,\boldsymbol{\Delta{\theta}}_r)\boldsymbol{\sigma}_\mathbf{x} 	{\mathbf{F}}(\boldsymbol{\Delta{\theta}}_t,\boldsymbol{\Delta{\theta}}_r)^H \right) \bigg),
	\end{aligned}
\end{equation}
$
		\mathcal{L}_{\boldsymbol{\omega}}(\boldsymbol{\omega}; \boldsymbol{\omega}^n)\triangleq\langle \ln p(\mathbf{s};\boldsymbol{\omega}) \rangle_{q(\mathbf{s}|\mathbf{y};\boldsymbol{\omega}^{n})}$ 
 and $K$ is a constant independent of  $\boldsymbol{\xi}$ and thus can be safely  omitted.
 \eqref{eq:1113_1} is derived from the evidence lower bound (ELBO) in the variational EM framework \cite{xu_joint_2024}.  Specifically, by taking the expectation of the complex Gaussian likelihood
\(\ln p(\mathbf{y}|\mathbf{x},\gamma;\boldsymbol{\Delta\theta}_t,\boldsymbol{\Delta\theta}_r)\)
and the Ising prior \(\ln p(\mathbf{s};\boldsymbol{\omega})\) with respect to the approximate posteriors \(q(\mathbf{x})\), \(q(\gamma)\), and \(q(\mathbf{s})\) obtained in the E-step, and by retaining only the terms that depend on $\boldsymbol{\xi}$, we can obtain  \eqref{eq:1113_1}, where all remaining terms independent of these parameters are absorbed into the constant \(K\).

%  \eqref{eq:1113_1} is derived from the evidence lower bound (ELBO) formulation in the variational EM framework \cite{xu_joint_2024}. 
%Specifically, by keeping  the term (i.e. $\mathbf{F}(\boldsymbol{\Delta{\theta}}_t,\boldsymbol{\Delta{\theta}}_r)$) that depend on 
%$\{\boldsymbol{\Delta\theta}_t,\boldsymbol{\Delta\theta}_r\}$, and $\boldsymbol{\omega}$,
%and calculating the expectation of the complex Gaussian likelihood 
%$\ln p(\mathbf{y}|\mathbf{x},\gamma;\boldsymbol{\Delta{\theta}}_t,\boldsymbol{\Delta{\theta}}_r) 
%\propto -\gamma\|\mathbf{y}-{\mathbf{F}}(\boldsymbol{\Delta{\theta}}_t,\boldsymbol{\Delta{\theta}}_r)\mathbf{x}\|^2$ and  $\ln p(\mathbf{\mathbf{s},\boldsymbol{\omega}})$
%with respect to the approximate posteriors $q(\mathbf{x})$, $q(\gamma)$ and $q(\mathbf{s})$ obtained in the E step, 
%we have 
%$\mathcal{L}_{\boldsymbol{\theta}}(\boldsymbol{\Delta{\theta}}_t,\boldsymbol{\Delta{\theta}}_r; \boldsymbol{\Delta{\theta}}^n_t,\boldsymbol{\Delta{\theta}}^n_r)+\mathcal{L}_{\boldsymbol{\omega}}(\boldsymbol{\omega}; \boldsymbol{\omega}^n)
%+ K$. 

1) \textit{Update of $\{\boldsymbol{\Delta\theta}_t,\boldsymbol{\Delta\theta}_r\}$}: From \eqref{eq:43}, we can see that  the trace term $\mathrm{Tr}\!\left( 	{\mathbf{F}}(\boldsymbol{\Delta{\theta}}_t,\boldsymbol{\Delta{\theta}}_r)\boldsymbol{\sigma}_\mathbf{x}	{\mathbf{F}}(\boldsymbol{\Delta{\theta}}_t,\boldsymbol{\Delta{\theta}}_r)^H \right)$ can theoretically improve the parameter $\{\boldsymbol{\Delta{\theta}}_t,\boldsymbol{\Delta{\theta}}_r\}$ estimation accuracy through the posterior covariance $\bm{\sigma}_x$, but it suffers from several drawbacks in practice. Specifically, under scenarios with low SNR or poor sparsity, the estimation of  $\boldsymbol{\sigma}_\mathbf{x}$ is often unstable, which may introduce significant 
numerical fluctuations and hinder the update of the angular offset vectors $\{\boldsymbol{\Delta{\theta}}_t,\boldsymbol{\Delta{\theta}}_r\}$. Moreover, compared to the dominant residual term $\|\mathbf{y}-{\mathbf{F}}(\boldsymbol{\Delta{\theta}}_t,\boldsymbol{\Delta{\theta}}_r)\bm{\mu}_x\|^2$, the numerical contribution of $\mathrm{Tr}\!\left( 	{\mathbf{F}}(\boldsymbol{\Delta{\theta}}_t,\boldsymbol{\Delta{\theta}}_r)\boldsymbol{\sigma}_\mathbf{x}	{\mathbf{F}}(\boldsymbol{\Delta{\theta}}_t,\boldsymbol{\Delta{\theta}}_r)^H \right)$
is marginal, while its gradient computation complexity increases substantially. Therefore, in the M-step updates of 
the proposed SF-TVBI algorithm, we omit this trace term and retain only the dominant residual term $\|\mathbf{y}-{\mathbf{F}}(\boldsymbol{\Delta{\theta}}_t,\boldsymbol{\Delta{\theta}}_r)\bm{\mu}_x\|^2$, which not only simplifies 
the derivation but also improves numerical stability. Accordingly, the optimization objective function in the M-step 
is given by
\begin{equation}\label{eq:27}
	\begin{aligned}
		&	\mathcal{L}^{(1)}_{\boldsymbol{\theta}}(\boldsymbol{\Delta{\theta}}_t,\boldsymbol{\Delta{\theta}}_r; \boldsymbol{\Delta{\theta}}^n_t,\boldsymbol{\Delta{\theta}}^n_r)
		\\&= - \langle \gamma \rangle \cdot 
		\left\| \mathbf{y} - {\mathbf{F}}(\boldsymbol{\Delta{\theta}}_t,\boldsymbol{\Delta{\theta}}_r)\boldsymbol{\mu}_\mathbf{x} \right\|^2.
	\end{aligned}
\end{equation}
Since \eqref{eq:27} is defined over the  entire angular grid offset $\{\bm{\Delta\theta}_t,\bm{\Delta\theta}_r\}$, a straightforward gradient evaluation must sweep all grid cells and is therefore computationally intensive. However, in multi-target sensing, only those cells with non-negligible target likelihood require updating. Therefore, we adopt a grid-selection strategy that updates only the angular grid offsets associated with cells deemed likely to contain targets at the current iteration.
Let $S_q$ denote the number of active elements (i.e., entries equal to one) 
in the estimated support vector $\hat{\mathbf{s}}$. 
$\boldsymbol{\Delta\hat{\theta}}_t \in \mathbb{C}^{M_tM_r\times S_q}$, 
$\boldsymbol{\Delta\hat{\theta}}_r \in \mathbb{C}^{M_tM_r\times S_q}$, 
and $\boldsymbol{\mu}_{\hat{\mathbf{x}}} \in \mathbb{C}^{S_q}$ can be constructed 
by collecting the transmit angular grid offset vectors $\boldsymbol{\Delta{\theta}}_t$, the receive angular grid offset vectors $\boldsymbol{\Delta{\theta}}_r$, 
and the mean values of the sparse signal $\boldsymbol{\mu}_\mathbf{x}$, respectively, at the $S_q$ active 
positions indicated by $\hat{\mathbf{s}}$. 
Moreover, in the   angular offset vectors $\{\boldsymbol{\Delta{\theta}}_t, \boldsymbol{\Delta{\theta}}_r\}$ update, only these $S_q$ grids corresponding 
to $\hat{\mathbf{s}}$ are required to be updated. Then, 
\eqref{eq:27} can be equivalently rewritten as
\begin{equation}\label{eq:30}
	\begin{aligned}
		&\mathcal{L}^{(2)}_{\boldsymbol{\theta}}(\boldsymbol{\Delta\hat{\theta}}_t,\boldsymbol{\Delta\hat{\theta}}_r; 
		\boldsymbol{\Delta\hat{\theta}}^n_t,\boldsymbol{\Delta\hat{\theta}}^n_r) \\
		&= - \langle \gamma \rangle \cdot 
		\left\| \mathbf{y} - {\mathbf{F}}(\boldsymbol{\Delta\hat{\theta}}_t, 
		\boldsymbol{\Delta\hat{\theta}}_r)\boldsymbol{\mu}_{\hat{x}} \right\|^2,
	\end{aligned}
\end{equation}
For notational simplicity, we drop the arguments of $\mathcal{L}^{(2)}_{\boldsymbol{\theta}}(\boldsymbol{\Delta\hat{\theta}}_t,\boldsymbol{\Delta\hat{\theta}}_r; 
\boldsymbol{\Delta\hat{\theta}}^n_t,\boldsymbol{\Delta\hat{\theta}}^n_r)$ and simply write $\mathcal{L}^{(2)}_{\boldsymbol{\theta}}$ hereafter.
Then,  the gradient ascent method is employed to maximize  $\mathcal{L}^{(2)}_{\boldsymbol{\theta}}$. 
The corresponding update expressions for $\boldsymbol{\Delta\hat{\theta}}_t$ and $\boldsymbol{\Delta\hat{\theta}}_r$ at the $(n+1)$-th iteration of the M-step in the SF-TVBI algorithm are given by
\begin{equation}\label{eq:0826_5}
	\boldsymbol{\Delta\hat{\theta}}^{n+1}_t = \boldsymbol{\Delta\hat{\theta}}^{n}_t + \iota_1^{n}\frac{\partial \mathcal{L}^{(2)}_{\boldsymbol{\theta}}}{\partial \boldsymbol{\Delta\hat{\theta}}_t}
\end{equation}
and
\begin{equation}\label{eq:0826_6}
	\boldsymbol{\Delta
		\hat{\theta}}^{n+1}_r = \boldsymbol{\Delta\hat{\theta}}^{n}_r + \iota_2^{n}\frac{\partial \mathcal{L}^{(2)}_{\boldsymbol{\theta}}}{\partial \boldsymbol{\Delta\hat{\theta}}_r},
\end{equation}
respectively, where $\iota_1^{n}$ and $\iota_2^{n}$ represent the step size determined by the
Armijo rule \cite{Armijo1966},  ${\partial \mathcal{L}^{(2)}_{\boldsymbol{\theta}}}/{\partial \boldsymbol{\Delta\hat{\theta}}_t}$ 
and ${\partial \mathcal{L}^{(2)}_{\boldsymbol{\theta}}}/{\partial \boldsymbol{\Delta\hat{\theta}}_r}$ 
are respectively expressed as
\begin{equation}\label{eq:28}
	\begin{aligned}
	\frac{\partial \mathcal{L}^{(2)}_{\boldsymbol{\theta}}}{\partial \boldsymbol{\Delta\hat{\theta}}_t}
		= &- \langle \gamma \rangle \cdot 
		\bigg( -2\Re \left\{ \boldsymbol{\mu}_{\hat{x}}^{H} \frac{\partial {\mathbf{F}}(\boldsymbol{\Delta\hat{\theta}}_t, 
			\boldsymbol{\Delta\hat{\theta}}_r)}{\partial\boldsymbol{\Delta\hat{\theta}}_t}^{H} \mathbf{y} \right\}
		\\&+ 2\Re \left\{  \boldsymbol{\mu}_{\hat{x}}^{H} \mathbf{F}(\boldsymbol{\Delta\hat{\theta}}_t, \boldsymbol{\Delta\hat{\theta}}_r)^{H} \frac{\partial {\mathbf{F}}(\boldsymbol{\Delta\hat{\theta}}_t, 
			\boldsymbol{\Delta\hat{\theta}}_r)}{\partial\boldsymbol{\Delta\hat{\theta}}_t}  \boldsymbol{\hat{\mu}}_x \right\} \bigg),
	\end{aligned}
\end{equation}
\begin{equation}\label{eq:29}
	\begin{aligned}
		\frac{\partial \mathcal{L}^{(2)}_{\boldsymbol{\theta}}}{\partial \boldsymbol{\Delta\hat{\theta}}_r}
		= &- \langle \gamma \rangle \cdot 
		\bigg( -2\Re \left\{ \boldsymbol{\mu}_{\hat{x}}^{H} \frac{\partial {\mathbf{F}}(\boldsymbol{\Delta\hat{\theta}}_t, 
			\boldsymbol{\Delta\hat{\theta}}_r)}{\partial\boldsymbol{\Delta\hat{\theta}}_r}^{H} \mathbf{y} \right\}
		\\&+ 2\Re \left\{ \boldsymbol{\mu}_{\hat{x}}^{H}  {\mathbf{F}}(\boldsymbol{\Delta\hat{\theta}}_t, 
		\boldsymbol{\Delta\hat{\theta}}_r)^{H} \frac{\partial {\mathbf{F}}(\boldsymbol{\Delta\hat{\theta}}_t, 
			\boldsymbol{\Delta\hat{\theta}}_r)}{\partial\boldsymbol{\Delta\hat{\theta}}_r}  \boldsymbol{\hat{\mu}}_x \right\} \bigg).
	\end{aligned}
\end{equation}

2) \textit{Update of $\{\boldsymbol{\omega}\}$}: Since the objective $\mathcal{L}_{\boldsymbol{\omega}}(\boldsymbol{\omega};\boldsymbol{\omega}^{(n)})$
does not admit a closed-form maximizer due to the presence of  $\ln z(\boldsymbol{\omega})$, we use pseudo-likelihood function to approximate the cross sparsity prior \cite{xu_joint_2024}, which is given by
\begin{equation}
	\mathrm{PL}(\mathbf{s};\boldsymbol{\omega})
	= \prod_{q\in\mathcal{Q}^2}
	p\big(s_q \mid \mathbf{s}_{\mathcal{N}_q};\boldsymbol{\omega}\big),
\end{equation}
where $\mathcal{N}_q$ denotes the index set of the neighbor nodes of $s_q$ in the factor graph induced by $p(\mathbf{s})$, and
$\mathbf{s}_{\mathcal{N}_q} \triangleq \{s_{q'} \mid q' \in \mathcal{N}_q\}$ collects the corresponding neighboring support variables.
For a binary Ising model as considered in this work, the local conditional distribution takes a
logistic form i.e.,
\begin{equation}
	\begin{aligned}
		p\big(s_q \mid \mathbf{s}_{\mathcal{N}_q};\boldsymbol{\omega}\big)
		&= \frac{\exp\big(s_q h_q\big)}
		{\exp(0\cdot h_q)+\exp(1\cdot h_q)}  \\
		&= \frac{\exp\big(s_q h_q\big)}{1+\exp\big(h_q\big)}.
	\end{aligned}
\end{equation}
where
\begin{equation}
	h_q \triangleq \sum_{q'\in\mathcal{N}_q}\omega_{q,q'}\,s_{q'}.
\end{equation}
Then, we can obtain the surrogate function as 
\begin{equation}
\widetilde{\mathcal{L}}_{\boldsymbol{\omega}}(\boldsymbol{\omega}; \boldsymbol{\omega}^n)=\langle \mathrm{PL}(\mathbf{s};\boldsymbol{\omega})\rangle_{q(\mathbf{s}|\mathbf{y};\boldsymbol{\omega}^{n})}
\end{equation}
The corresponding update expressions for $\boldsymbol{\omega}$  at the $(n+1)$-th iteration of the M-step in the SF-TVBI algorithm are given by
\begin{equation}\label{eq:x}
	\bm{\omega}^{n+1}=\bm{\omega}^{n}+\iota^n_3\frac{\partial {\widetilde{\mathcal{L}}_{\boldsymbol{\omega}}(\boldsymbol{\omega}; \boldsymbol{\omega}^n)}}{\partial \boldsymbol{\omega}},
\end{equation}
where $\iota^n_3$ represents the step size determined by the
Armijo rule \cite{Armijo1966}. The detailed derivation is similar to that in [27] and is omitted here for brevity.

Algorithm~1 summarizes the proposed SF-TVBI procedure. As for computational complexity, we can see that the  per--inner-iteration cost of the
SF-TVBI E-step is $\mathcal{O}(Q_1^{3})$, dominated by the matrix inversion operation in
\eqref{eq:15}. The per-inner-iteration cost of the M-step is
$\mathcal{O}(M_t M_r S_q)$, primarily due to the gradient evaluations in
\eqref{eq:28} and \eqref{eq:29}. Hence, the overall complexity of SF-TVBI is
$
\mathcal{O}\!\left(
I_{\mathrm{out},1} I_{\mathrm{in},1} Q_1^{3}
+
I_{\mathrm{out},1} I_{\mathrm{in},2} M_t M_r S_q
\right),
$ where $I_{\mathrm{out},1}$, $I_{\mathrm{in},1}$, and $I_{\mathrm{in},2}$ denote the number of outer-loop iterations, the number of E-step inner-loop iterations, and the number of M-step inner-loop iterations, respectively. 
For comparison, the Turbo-VBI algorithm entails
$
\mathcal{O}\!\left(
I_{\mathrm{out},2} I_{\mathrm{in},1} Q_1^{3}
+
I_{\mathrm{out},2} M_t M_r Q_1
\right)
$, where $I_{\mathrm{out},2}$ denotes the number of outer-loop iterations of the Turbo-VBI algorithm.
In practical deployments, since $I_{\mathrm{out},2}\gg I_{\mathrm{out},1}$
and $Q_1>S_q$, SF-TVBI incurs a substantially lower computational burden
than Turbo-VBI.
The computational complexity of the proposed SF-TVBI algorithm can be further reduced by incorporating the idea of subspace constrained variational Bayesian inference (SC-VBI)  \cite{10852191}, in which case the resulting complexity becomes $\mathcal{O}\!\left(
I_{\mathrm{out},1} I_{\mathrm{in},1} Q_1 M_r L
+
I_{\mathrm{out},1} I_{\mathrm{in},2} M_t M_r S_q
\right)$.
\begin{algorithm}[t]
	\caption{SF-TVBI Algorithm}
	\KwIn{$\mathbf{y}$, $\boldsymbol{\Delta\theta}^1_t=0$, $\boldsymbol{\Delta\theta}^1_r=0$, $\boldsymbol{\theta}_t$, $\boldsymbol{\theta}_r$,  $I_{out,1}$,  $I_{in,1}$, $I_{in,2}$, $a_{q}$, $b_{q}$, $\bar{a}_{q}$, $\bar{b}_{q}$, $\forall q \in \mathcal{Q}^2$, $c$, $d$ hyperparameters of corss sparsity $\boldsymbol{\omega}$, $\tau_s$, $\tau$}
	\KwOut{$\boldsymbol{\Delta\theta}_t$, $\boldsymbol{\Delta\theta}_r$, $q(\mathbf{s})$}
	
	\For{$i_{out} = 1,\dots,I_{out,1}$}{
		\textbf{SF-TVBI-E step}\\
		\textbf{\%Module A :VBI Estimator}\\
		\While{not converged and $i_{in,1}\leq I_{in,1}$}{
			Update  $q(\mathbf{x})$, $q(\boldsymbol{\rho})$, $q(\boldsymbol{s})$ and $q(\gamma)$ using  \eqref{eq:k14}, \eqref{eq:1003_1}, \eqref{eq:k0826_3},\eqref{eq:1031_1} and \eqref{eq:1031_2}, respectively;
		}
		\textbf{\%Module B: Message Passing in cross sparsity}\\
		Calculate the output messages $g^B_{q}(s_{q})$.\\
		Perform sum-product message  passing over the support factor graph and calculate the output messages according to Appendix \ref{A}.\\
		\textbf{SF-TVBI-M step}\\
		Calculate the estimate vector $\hat{\mathbf{s}}$, using \eqref{eq:0904_1}.\\
			Obtain  $\boldsymbol{\Delta\hat{\theta}}^{1}_t$ and $\boldsymbol{\Delta\hat{\theta}}^{1}_r$ according to  $\mathbf{\hat{s}}$, $\boldsymbol{\Delta\theta}^{i_{out}}_t$ and $\boldsymbol{\Delta\theta}^{i_{out}}_r$.\\
		\While{$i_{in,2}\leq I_{in,2}$}{
		Update $\boldsymbol{\Delta\hat{\theta}}^{i_{in,2}+1}_t$, using \eqref{eq:28} and \eqref{eq:0826_5}.\\ Update $\boldsymbol{\Delta\hat{\theta}}^{i_{in,2}+1}_r$, using \eqref{eq:29} and \eqref{eq:0826_6}.\\Update $\boldsymbol{\omega}^{i_{in,2}+1}$, using \eqref{eq:x}.\\
			\If{$\|\boldsymbol{\Delta\hat{\theta}}^{i_{in,2}+1}_t-\boldsymbol{\Delta\hat{\theta}}^{i_{in,2}}_t\|+\|\boldsymbol{\Delta\hat{\theta}}^{i_{in,2}+1}_r-\boldsymbol{\Delta\hat{\theta}}^{i_{in,2}}_r\|+\|\boldsymbol{\omega}^{i_{in,2}+1}-\boldsymbol{\omega}^{i_{in,2}}\|\leq \tau$}{
				\textbf{break}\;
			}
			$i_{in,2}\leftarrow i_{in,2}+1$.
		}
		Update $\boldsymbol{\Delta\theta}^{i_{out}+1}_{t}$ and $\boldsymbol{\Delta\theta}^{i_{out}+1}_{r}$, by using $\boldsymbol{\Delta\hat{\theta}}^{i_{in,2}}_t$ and  $\boldsymbol{\Delta\hat{\theta}}^{i_{in,2}}_t$, respectively.\\
		$i_{out,1}\leftarrow i_{out,1}+1$.
	}
\end{algorithm}

\section{Numerical Results}
In this section, we present numerical results to validate the effectiveness of the proposed cross sparsity model and the SF-TVBI algorithm. We consider a colocated MIMO radar with $M_t=16$ transmit and $M_r=16$ receive antennas operating at a carrier frequency of $4\,\text{GHz}$, where the inter-element spacing in both arrays is set to $\lambda/2$, and the angular domain is discretized into $Q=16$ grids. Based on \cite{chen_joint_2024}, the large-scale fading models for the direct paths,
$g_k$, $k \in \mathcal{K}$, and for the first-order paths,
$h_{i,j}$, $i,j \in \mathcal{K},\, i \neq j$, can be expressed as
\begin{equation}\label{eq:1003_2}
	g_k=\sqrt{\frac{\lambda^2 \kappa_k}{64\pi^3 d_k^4}}
\end{equation}
and
\begin{equation}\label{eq:1003_3}
	h_{i,j}=\sqrt{\frac{\lambda^2\,\Psi_{i,j}\,\Psi_{j,i}}{(4\pi)^4\, d_i^2\, l_{i,j}^2\, d_j^2}},
\end{equation}
respectively.
Here, $\kappa_k$ denotes the radar cross section (RCS) of the $k$-th target along the direct path and is set to $-10\,\text{dBsm}$, while $\Psi_{i,j}$ denotes the bistatic RCS of the $i$-th target along the first-order path in the look  of the $j$-th target (i.e., scattering from the $i$-th target toward the $j$-th target) and is set to $0\,\text{dBsm}$ \cite{c10176877}.  $d_k$ and $l_{i,j}$ denote the distance between the $k$-th target and the radar, and the distance between the $i$-th and $j$-th targets, respectively.
 The initial value for $\boldsymbol{\omega}$ is empirically determined, and then updated by the M-step of the SF-TVBI. The transmit power of the MIMO radar is set as $P_T=30\,\text{dBm}$.  We assume  that
the direct angles of targets are randomly distributed within the interval $[-\frac{\pi}{2}, +\frac{\pi}{2}]$, with the constraint that they lie on distinct angular grid points.  Algorithm 1 uses the following default parameters: $I_{out,1}=20$, $I_{in,1}=10$, $I_{in,2}=20$, $\tau_s=0.8$ and $\tau_s=0.8$.  For comparison, we consider the  OMP algorithm, the CS
method in continuous domain (CSCD)  \cite{zheng_detection_2024} and the Turbo-VBI algorithm  \cite{liu_robust_2020} as  benchmarks. The estimation accuracy is evaluated in terms of the root mean-squared error (RMSE), defined as
\begin{equation}
	\mathrm{RMSE} = \sqrt{ \frac{1}{K} \sum_{k=1}^{K} 
		\frac{1}{J} \sum_{j=1}^{J} 
		\left( \theta_k^{(j)} - \hat{\theta}_k^{(j)} \right)^2 },
\end{equation}
where $\theta_k^{(j)}$ and $\hat{\theta}_k^{(j)}$ denote the true and estimated direct-path angles of the $k$-th target in the $j$-th trial, respectively, and $J=500$ independent Monte Carlo trails  are performed.

\begin{figure}[t]
	\centering
	\includegraphics[width=2.7in]{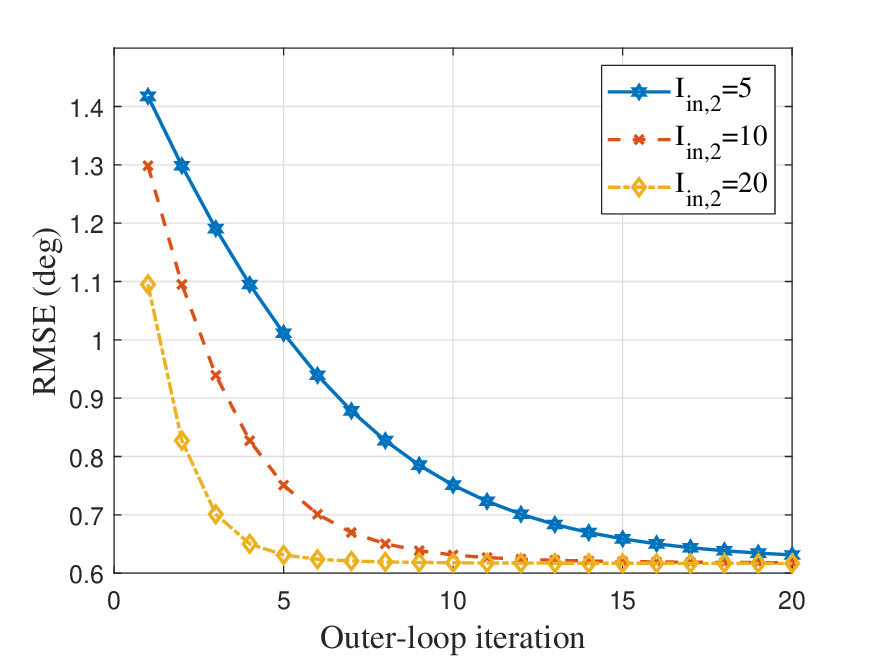}
	\caption{RMSE  versus the out-loop iteration number.}
	\label{fig10003_1}
\end{figure}
\begin{figure}[t]
	\centering
	\includegraphics[width=3.8in]{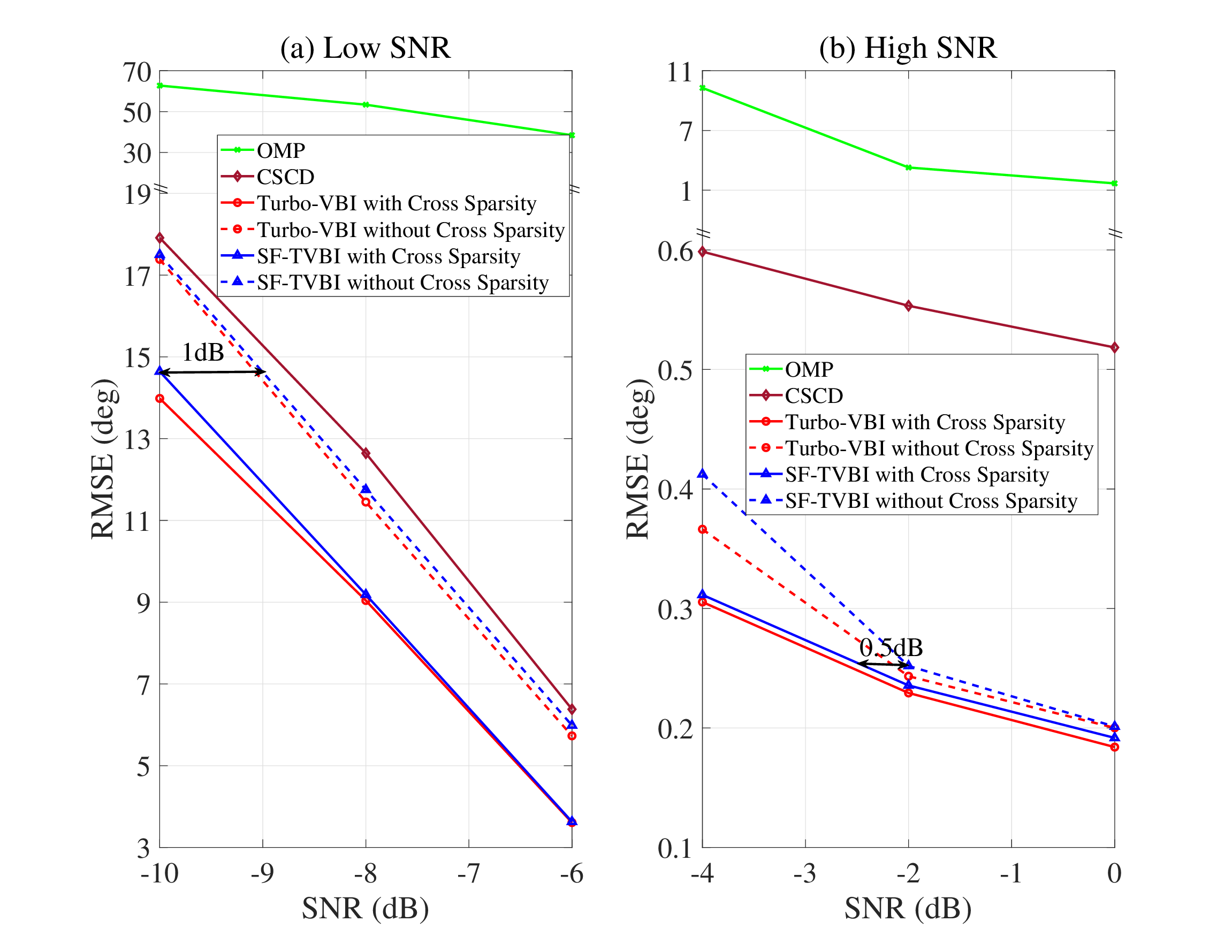}
	\caption{RMSE  versus SNR.}
	\label{fig10003_2}
\end{figure}
\begin{figure}[t]
	\centering
	\includegraphics[width=2.7in]{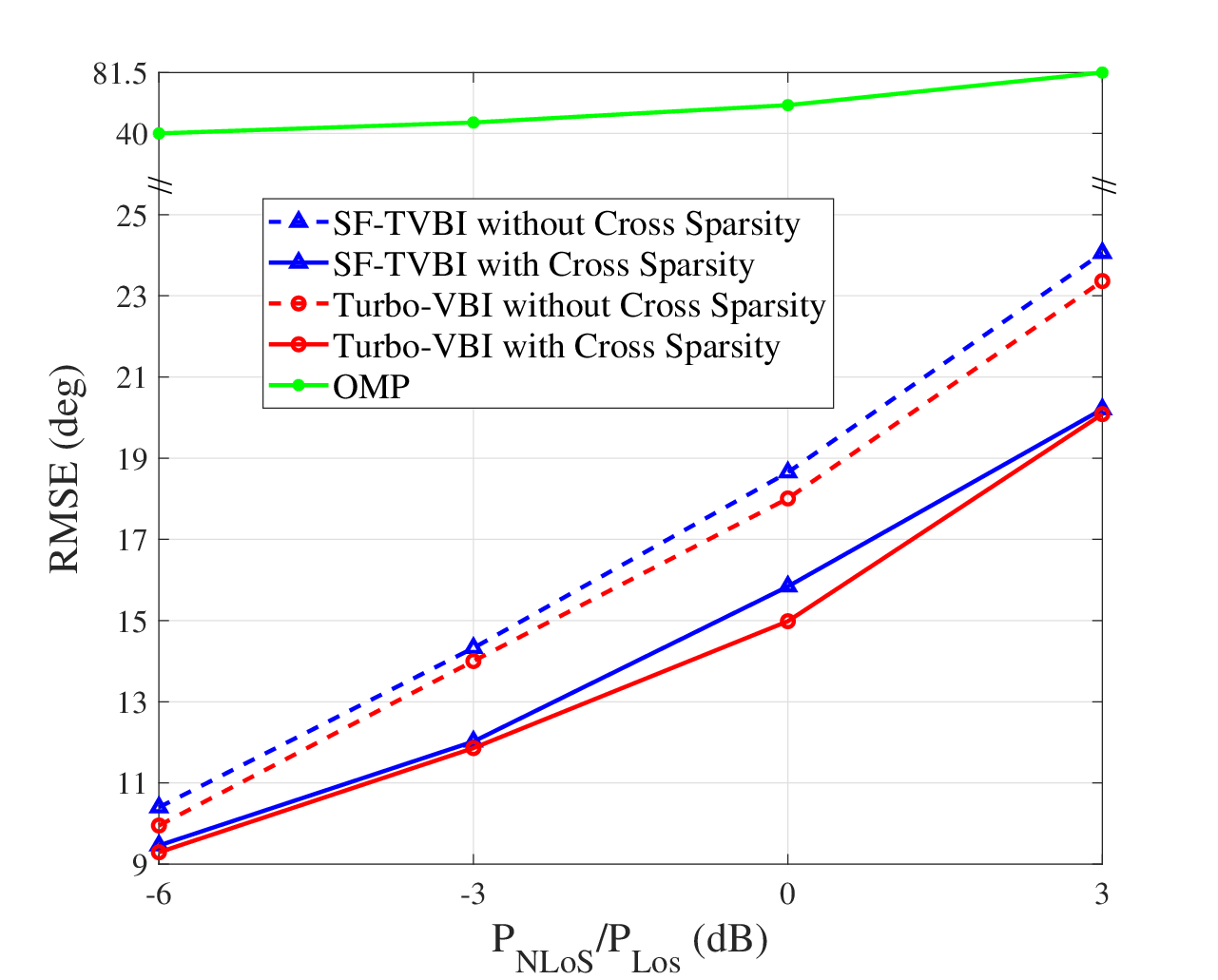}
	\caption{ RMSE versus $\mathrm{P_{NLos}}/\mathrm{P_{Los}}$.}
	\label{fig10003_3}
\end{figure}
\begin{figure}[t]
	\centering
	\includegraphics[width=2.7in]{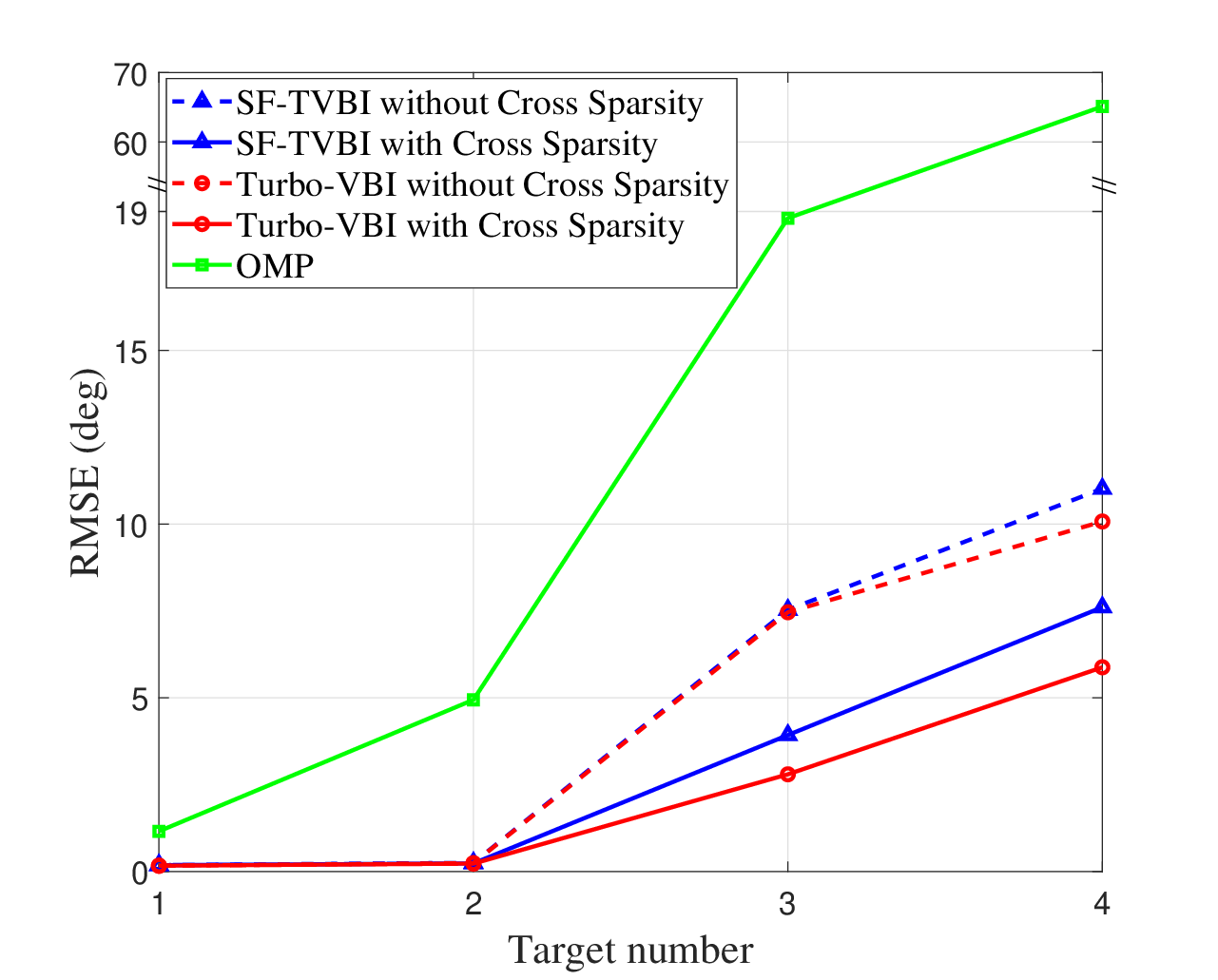}
	\caption{ RMSE versus target number.}
	\label{fig10003_4}
\end{figure}
\begin{figure}[t]
	\centering
	\includegraphics[width=2.7in]{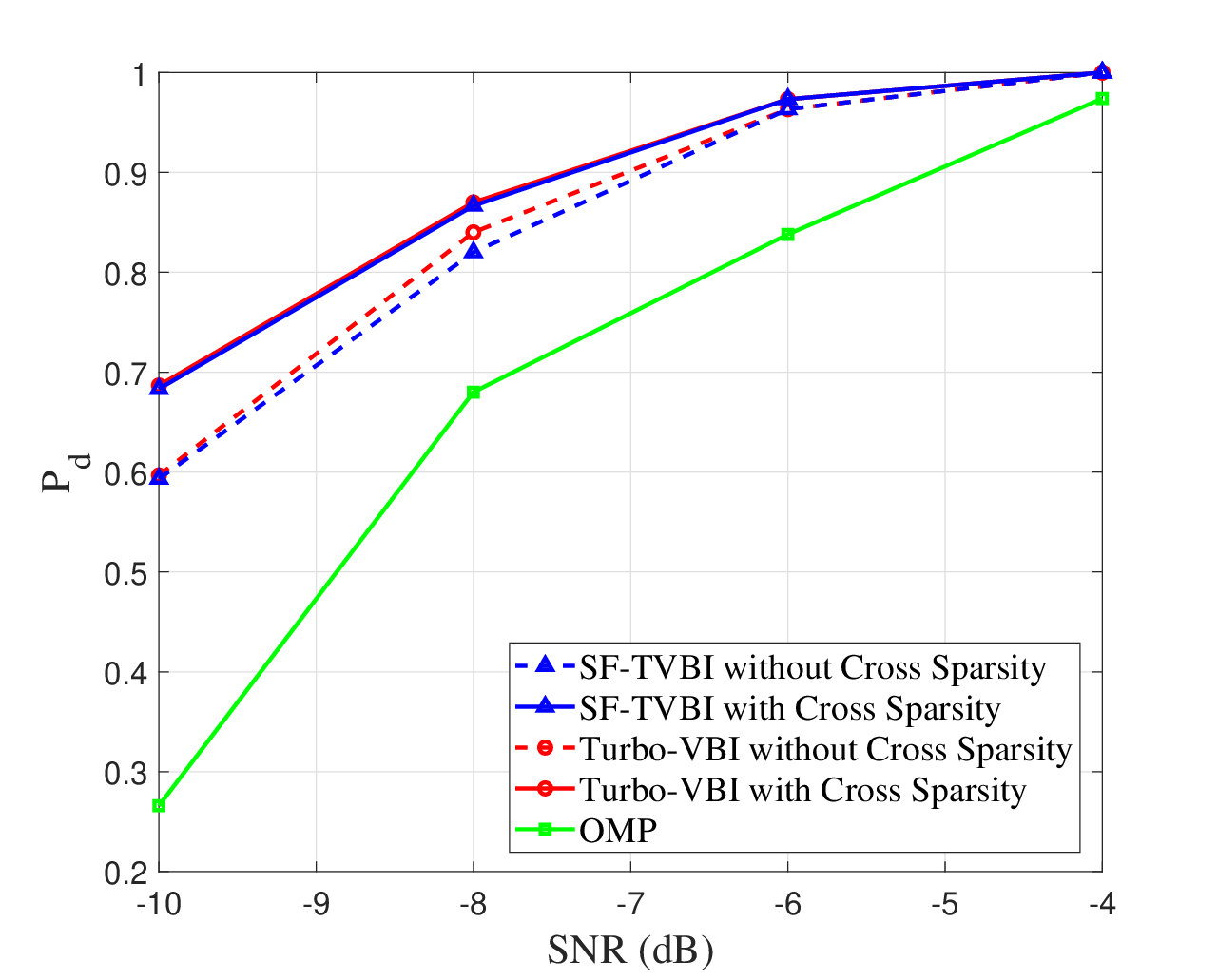}
	\caption{ Detect probability versus SNR.}
	\label{fig10026_1}
	\vspace{-3mm}
\end{figure}

Before proceeding to the performance comparison, we first investigate the convergence behavior of the proposed SF-TVBI algorithm with cross sparsity, as illustrated in Fig.~\ref{fig10003_1}. It can be observed that as the number of inner-loop M-step iterations increases, the required number of outer-loop iterations decreases accordingly. This indicates that fewer but more effective E-step updates are sufficient to ensure algorithmic convergence. Therefore, the proposed SF-TVBI algorithm achieves a lower overall computational complexity.

In Fig. \ref{fig10003_2}, we investigate the RMSE performance achieved by the considered algorithms  under different SNR conditions.  As can be seen,   the Bayesian inference-based methods (SFTVBI
and Turbo-VBI) and the CSCD algorithm significantly outperform the conventional OMP algorithm. This performance gain is achieved by separately estimating the DoD and DoA of the first-order scattering paths and the direct (LoS) path, thereby substantially mitigating the interference of the first-order components on the direct-path estimation. Moreover, the  Bayesian inference-based methods  consistently achieves better estimation accuracy than CSCD, especially in the high-SNR regime, owing to its more effective exploitation of the available prior information. In addition, by leveraging the proposed cross sparsity structure, the estimation performance can be further enhanced by approximately $1\,\text{dB}$ at low SNRs and $0.5\,\text{dB}$ at higher SNRs, which demonstrates the effectiveness of cross sparsity in improving multi-target estimation. Finally, it can be observed that the proposed SF-TVBI algorithm attains estimation accuracy comparable to that of the Turbo-VBI algorithm, thereby confirming its ability to deliver near-identical performance with a substantially reduced computational burden.

Fig.~\ref{fig10003_3} shows the RMSE versus the power ratio of the first-order to direct paths i.e., $\mathrm{P_{NLoS}}/\mathrm{P_{LoS}}$, where 
$\mathrm{P}_{\mathrm{LoS}}$ and 
$\mathrm{P}_{\mathrm{NLoS}}$
denote the received powers of the direct and first-order paths, respectively, and the ratio is varied by changing the distances between the targets and the radar. It can be observed that the RMSE reduction achieved by exploiting the cross sparsity structure increases from about $10\%$, when first-order paths are weak (e.g., $\mathrm{P_{NLoS}}/\mathrm{P_{LoS}}<-3\,\mathrm{dB}$), to roughly $20\%$ as they strengthen (e.g., $\mathrm{P_{NLoS}}/\mathrm{P_{LoS}}\geq-3\,\mathrm{dB}$), demonstrating a growing performance enhancement as $\mathrm{P_{NLoS}}/\mathrm{P_{LoS}}\geq-3\,\mathrm{dB}$ increases. Overall, as the energy of the first-order paths increases, a more significant RMSE reduction is achieved. This performance gain mainly stems from the improved posterior probability estimation of the grid support vector $\mathbf{s}$ enabled by the cross sparsity prior.

In Fig. \ref{fig10003_4}, we investigate the RMSE performance of the considered methods  under the different target numbers, where we  set the receive SNR at $\mathrm{SNR}=-2\mathrm{dB}$. As shown, the performance gain achieved by the proposed cross sparsity structure increases with the number of targets. This improvement mainly results from two aspects. First, as the number of targets increases, the number of reflection paths among them also grows, and this increase can be effectively leveraged by the cross sparsity structure to exploit transmit-receive path correlations and improve estimation accuracy. Second, under a fixed receive SNR, the increase in the number of paths reduces the per-path SNR, while the cross sparsity structure mitigates this resulting performance degradation by fully utilizing the first-order reflection paths. 
%Furthermore, it can be observed that the performance gain brought by the cross sparsity structure does not further increase as the number of targets continues to grow, which appears counterintuitive. This phenomenon mainly occurs because a larger number of targets introduces more propagation paths, and under a fixed number of radar antennas, the increased number of paths leads to higher angular ambiguity. The resulting ambiguity degrades the system’s ability to distinguish closely spaced paths, thereby limiting the performance gain provided by the cross sparsity structure.

In Fig. \ref{fig10026_1}, we evaluate the target detection probability $P_{d}$ of the proposed algorithm across a range of SNR levels, where $P_d$ is  defined as the probability of  targets being identified in the correct grid cell. As can be seen,  the OMP algorithm exhibits a significantly inferior detection probability compared to the VBI-based methods at the low SNR regime This demonstrates that the incorporation of prior information in the VBI framework effectively enhances the target grid detection capability. Furthermore, the VBI-based algorithm with the proposed cross sparsity structure achieves  higher detection probability than its counterpart without cross sparsity, and the improvement becomes more pronounced as the SNR decreases. This is because the cross sparsity model captures the inherent dependencies among multiple targets in multipath environments, and the useful information from multiple paths helps compensate for the weak individual signals at low SNR, resulting in more robust detection. In addition, the proposed SF-TVBI algorithm attains detection performance very close to that of the Turbo-VBI algorithm, confirming its effectiveness.

\section{Conclusion}
In this paper, we studied a multi-target system in a multipath environment, where inter-target reflections give rise to first-order paths. By exploiting the geometric property that first-order paths share their AoAs/AoDs with the direct paths of other targets, we proposed a novel cross sparsity structure under a 3LHS probability model to enhance sensing accuracy. In addition, to reduce the computational complexity of Turbo-VBI, we proposed a SF-TVBI algorithm, which integrates an efficient message-passing strategy to enable tractable probabilistic information exchange within the cross sparsity factorization and a two-timescale update scheme that reduces the update frequency of the high-dimensional sparse vector. Simulation results validated the effectiveness of the proposed cross sparsity  and showed that SF-TVBI attains performance comparable to Turbo-VBI while requiring substantially lower computational complexity.

{\appendix
\subsection{Derivation of Approximate Marginal Posteriors}	\label{B}
% ---------- Appendix A: Variational Posterior Derivations ----------
We begin by isolating the terms that involve $\mathbf{x}$ while treating $(\boldsymbol{\theta}_t,\boldsymbol{\theta}_r)$ as deterministic and taking expectations with respect to the remaining factors. Then, $q(\mathbf{x})=\mathcal{CN}(\mathbf{x};\boldsymbol{\mu}_\mathbf{x},\boldsymbol{\sigma}_\mathbf{x} )$ can be derived as
\begin{equation}
	\small
	\begin{aligned}
		\ln q(\mathbf{x}) 
		&\propto \left\langle \ln p(\mathbf{y}, \mathbf{x}, \boldsymbol{\rho}, \gamma; 
		\boldsymbol{\Delta{\theta}}_t,\boldsymbol{\Delta{\theta}}_r) 
		\right\rangle_{q(\boldsymbol{\rho}) q(\gamma)} \\[2pt]
		&\propto \left\langle \ln p(\mathbf{y} | \mathbf{x}, \gamma; 
		\boldsymbol{\Delta{\theta}}_t,\boldsymbol{\Delta{\theta}}_r) \right\rangle_{q(\gamma)} 
		\\&\quad+ \left\langle \ln p(\mathbf{x} | \boldsymbol{\rho}) \right\rangle_{q(\boldsymbol{\rho})} \\[2pt]
		&\propto - \langle \gamma \rangle \left\| \mathbf{y} - 
		{\mathbf{F}}(\boldsymbol{\Delta{\theta}}_t,\boldsymbol{\Delta{\theta}}_r)\mathbf{x} \right\|^2 
		\\&\quad- \mathbf{x}^H \, \mathrm{diag}\!\left(\langle \boldsymbol{\rho} \rangle \right)\mathbf{x} \\[2pt]
		&\propto - \mathbf{x}^H \boldsymbol{\sigma}_\mathbf{x}^{-1} \mathbf{x} 
		+ 2 \Re \Big\{ \mathbf{x}^H \boldsymbol{\sigma}_\mathbf{x}^{-1} \boldsymbol{\mu}_\mathbf{x} \Big\},
	\end{aligned}
	\label{eq:14}
\end{equation}
Next, $
q(\boldsymbol{\rho}) = \prod_{n=1}^{Q_1} \mathrm{Ga}(\rho_{n};\,\tilde{a}_{n}, \tilde{b}_{n})
$ can be derived as
\begin{equation}
	\small
	\begin{aligned}
		\ln q(\boldsymbol{\rho}) 
		&\propto \Big\langle \ln p(\mathbf{y}, \mathbf{x}, \boldsymbol{\rho}, \mathbf{s}, \gamma)
		\Big\rangle_{q(\mathbf{x}) q(\mathbf{s}) q(\gamma)} \\[2pt]
			&\propto \Big\langle \ln p(\mathbf{x} | \boldsymbol{\rho}) \Big\rangle_{q(\mathbf{x})} 
		+ \Big\langle \ln p(\boldsymbol{\rho} | \mathbf{s}) \Big\rangle_{q(\mathbf{s})} \\[2pt]
		&\propto \sum_{n=1}^{Q_1} \ln \rho_{n}
		- \sum_{n=1}^{Q_1} \rho_{n} \big\langle x_{n}^{2}\big\rangle \\
		&\quad + \sum_{n=1}^{Q_1} \langle s_{n} \rangle \big[(a_{n}-1)\ln \rho_n -  b_{n}\rho_n \big] \\
		&\quad + \sum_{n=1}^{Q_1} \langle 1-s_{n} \rangle \big[(\bar{a}_{n}-1)\ln \rho_n - \bar{a}_{n} \rho_{n} \big] \\
		&\propto \sum_{n=1}^{Q_1} \big[ \langle s_{n} \rangle a_{n} + \langle 1-s_n \rangle \bar{a}_{n} \big] \ln \rho_{n} \\
		&\quad - \sum_{n=1}^{Q_1} \big[ \langle s_{n} \rangle b_{n} + \langle 1-s_n \rangle \bar{a}_{n} + \langle x_{n}^{2}\rangle \big] \rho_{n} \\
		&\propto \sum_{n=1}^{Q_1} (\tilde{a}_{n} - 1)\ln \rho_{n} 
		- \sum_{n=1}^{Q_1} \tilde{b}_{n} \rho_{n}.
	\end{aligned}
	\label{eq:rho_update}
\end{equation}
Then, $q(\mathbf{s}) = \prod_{n=1}^{Q_1} \tilde{\lambda}_n^{\,s_{n}} \left(1-\tilde{\lambda}_n\right)^{\,1-s_{n}}$ can be derived as
\begin{equation}\label{eq:qs_update}
	\small
	\begin{aligned}
		\ln q(\mathbf{s})
		&\propto \big\langle \ln p(\mathbf{y}, \mathbf{x}, \boldsymbol{\rho}, \mathbf{s}, \gamma) \big\rangle_{q(\mathbf{x})\,q(\boldsymbol{\rho})\,q(\gamma)} \\[2pt]
		&\propto \big\langle \ln p(\boldsymbol{\rho}|\mathbf{s}) \big\rangle_{q(\boldsymbol{\rho})} + \ln p(\mathbf{s}) \\
		&\propto \sum_{n=1}^{Q_1} (1-s_{n})\,\ln\!\big[(1-\pi_{n})\bar{C}_{n}\big] \\
		&\quad+\sum_{n=1}^{Q_1} s_{n} \ln(\pi_{n} C_{n})\\
		&\propto \sum_{n=1}^{Q_1} s_{n} \ln \tilde{\lambda}_n
		+ \sum_{n=1}^{Q_1} (1-s_{n})\,\ln(1-\tilde{\lambda}_n).
	\end{aligned}
\end{equation}
Finally, $
\small
q(\gamma) = \mathrm{Ga}(\gamma; \tilde{c}, \tilde{d})
$ can be derived as
\begin{equation}\label{eq:qgamma}
	\small
	\begin{aligned}
		\ln q(\gamma) 
		&\propto \left\langle \ln  p(\mathbf{y}, \mathbf{x}, \boldsymbol{\rho}, \mathbf{s}, \gamma)  \right\rangle_{q(\mathbf{x})q(\boldsymbol{\rho})q(\mathbf{s})q(\gamma)} + \ln p(\gamma) \\[2pt]
		&\propto \left\langle \ln p(\mathbf{y}|\mathbf{x},\gamma;
		\boldsymbol{\Delta{\theta}}_t,\boldsymbol{\Delta{\theta}}_r) \right\rangle_{q(\mathbf{x})} + \ln p(\gamma) \\
		&\propto Q_1 \ln \gamma
		- \gamma \left\langle \|\mathbf{y} - 	{\mathbf{F}}(\boldsymbol{\Delta{\theta}}_t,\boldsymbol{\Delta{\theta}}_r) \mathbf{x}\|^2 \right\rangle_{q(\mathbf{x})}  \\
		&\quad + (c-1)\ln \gamma - d\,\gamma \\
		&\propto (Q_1+c-1)\ln \gamma\\
		&\quad- \Big(d + \left\langle \|\mathbf{y} - 	{\mathbf{F}}(\boldsymbol{\Delta{\theta}}_t,\boldsymbol{\Delta{\theta}}_r) \mathbf{x}\|^2 \right\rangle_{q(\mathbf{x})} \Big)\gamma\\
		&\propto (\tilde{c}-1)\ln \gamma - \tilde{d}\gamma.
	\end{aligned}
\end{equation}
\subsection{Message Update Equations for Module B}\label{A}
1) \textit{Message Passing Over the Path} $g^B_{r,(i-1)Q+j} \rightarrow s_{(i-1)Q+j},i,j\in\{1,2,\cdots,Q\}$: The factor node $g^B_{r,(i-1)Q+j}$ represent the output message from Module A, which is given by
\begin{equation}\label{eq:0828_3}
	\small
	\begin{aligned}
		g^B_{r,(i-1)Q+j} &= \frac{q(s_{r,(i-1)Q+j})}{g^{A}_{r,(i-1)Q+j}(s_{r,(i-1)Q+j})}\\
		&\propto \pi^B_{(i-1)Q+j}\delta(s_{r,(i-1)Q+j}-1)\\&\quad+(1-\pi^B_{(i-1)Q+j})\delta(s_{r,(i-1)Q+j}),
	\end{aligned}
\end{equation}
where  $\pi^B_{q}=\frac{\tilde{\lambda}_q(1-\pi_{q})}{\tilde{\lambda}_q(1-\pi_{q})+(1-\tilde{\lambda}_q)\pi_{q}}$ with $\pi_{q}$ being the probability of $s_{q}=1$ in the output message $g^A_{r,q}(s_{q})$ from Module B.

2) Message Passing Over the Path $s_{(i-1)Q+j}\rightarrow \varphi_{\{i,j\}}^{\{i,i\}}$: We denote the message passing over this path as $\nu_{s_{(i-1)Q+j}\rightarrow \varphi_{\{i,j\}}^{\{i,i\}}}$, which is given by
\begin{equation}\label{eq:0829_3}
	\small
	\begin{aligned}
		&	\nu_{ s_{(i-1)Q+j}\rightarrow\varphi_{\{i,j\}}^{\{i,i\}} } \\&\propto g^B_{r,(i-1)Q+j}\times \!\!\nu_{ \varphi_{\{i,j\}}^{\{j,j\}}\rightarrow s_{(i-1)Q+j}} \\&\propto
		\pi^{p,\{i,j\}}_{tp1}\delta(s_{(i-1)+j}-1)+(1-	\pi^{p,\{i,j\}}_{tp1})\delta(s_{(i-1)+j}),
	\end{aligned}
\end{equation}
where 
\begin{equation}\label{eq:0829_4}
	\small
	\begin{aligned}
		&\pi^{p,\{i,j\}}_{tp1} \\&\triangleq
		\frac{\pi^{n,\{i,j\}}_{tp2}\,\pi^B_{(i-1)Q+j}}
		{\pi^{n,\{i,j\}}_{tp2}\,\pi^B_{(i-1)Q+j} + \big(1-\pi^{n,\{i,j\}}_{tp2}\big)\big(1-\pi^B_{(i-1)Q+j}\big)}
	\end{aligned}
\end{equation}
represents the probability of $s_{(i-1)Q+j}=1$ in the message passing over the path ${s_{(i-1)Q+j}\rightarrow \varphi_{\{i,j\}}^{\{i,i\}}}$, and $\pi^{n,\{i,j\}}_{tp2}$ represents the probability of $s_{(i-1)Q+j}=1$ in the message passing over the path ${\varphi_{\{i,j\}}^{\{j,j\}}\rightarrow s_{(i-1)Q+j}}$,   whose detailed expression will be specified later.

3) Message Passing Over the Path $\varphi_{\{i,j\}}^{\{i,i\}}\rightarrow s_{(i-1)Q+i}$: We denote the message passing over this path as $\nu_{\varphi_{\{i,j\}}^{\{i,i\}}\rightarrow s_{(i-1)Q+i}}$, which is given by
\begin{equation}\label{eq:0829_1}
	\small
	\begin{aligned}
		&\nu_{\varphi_{\{i,j\}}^{\{i,i\}}\rightarrow s_{(i-1)Q+i}} \\&\propto	\sum_{s_{(i-1)Q+j}}\!\!\nu_{ s_{(i-1)Q+j}\rightarrow\varphi_{\{i,k\}}^{\{i,i\}}} \times\varphi(s_{(i-1)Q+j},s_{(i-1)Q+i})\\&\propto
		\pi^{p,\{i,j\}}_{tr1}\delta(s_{(i-1)+i}-1)+(1-	\pi^{p,\{i,j\}}_{tr1})\delta(s_{(i-1)+i}),
	\end{aligned}
\end{equation}
where
\begin{equation}
	\small
	\pi^{p,\{i,j\}}_{tr1}\triangleq
	\frac{\pi^{p,\{i,j\}}_{tp1} e^{\omega} + \left(1-\pi^{p,\{i,j\}}_{tp1}\right)}
	{\pi^{p,\{i,j\}}_{tp1} e^{\omega} + \pi^{p,\{i,j\}}_{tp1} + 2\left(1-\pi^{p,\{i,j\}}_{tp1}\right)},
\end{equation}
represents the probability of $s_{(i-1)Q+i}=1$ in the message passing over the path $\varphi_{\{i,j\}}^{\{i,i\}}\rightarrow s_{(i-1)Q+i}$.

4) Message Passing Over the Path $ s_{(i-1)Q+i}\rightarrow\varphi_{\{i,j\}}^{\{i,i\}}$: We denote the message passing over this path as $\nu_{ s_{(i-1)Q+i}\rightarrow\varphi_{\{i,j\}}^{\{i,i\}}}$, which is given by
\begin{equation}\label{eq:0829_7}
	\small
	\begin{aligned}
		&\nu_{s_{(i-1)Q+i}\rightarrow\varphi_{\{i,j\}}^{\{i,i\}}} \\&\propto\!\!\!
		\prod_{k=1,k\neq j}^{Q}\!\! \nu_{\varphi_{\{i,k\}}^{\{i,i\}}\rightarrow s_{(i-1)Q+i}} \times  \nu_{\varphi_{\{k,i\}}^{\{i,i\}}\rightarrow s_{(i-1)Q+i}} \times g^B_{r,(i-1)Q+i}\\&\propto \pi^{n,\{i,j\}}_{tr1} \delta(s_{(j-1)+j}-1)+(1-	\pi^{n,\{i,j\}}_{tr1} )\delta(s_{(j-1)+j}),
	\end{aligned}
\end{equation}
where
\begin{equation}
	\small
	\pi^{n,\{i,j\}}_{tr1} \triangleq
	\frac{v(1)_{s_{(i-1)Q+i}\rightarrow\varphi_{\{i,j\}}^{\{i,i\}}}}
	{v(1)_{s_{(i-1)Q+i}\rightarrow\varphi_{\{i,j\}}^{\{i,i\}}}+v(0)_{s_{(i-1)Q+i}\rightarrow\varphi_{\{i,j\}}^{\{i,i\}}}},
\end{equation}
represents the probability of $s_{(i-1)Q+i}=1$ in the message passing over the path $s_{(i-1)Q+i}\rightarrow\varphi_{\{i,j\}}^{\{i,i\}}$, and we denote that
\begin{equation}
	\small
	v{(1)}_{s_{(i-1)Q+i} \rightarrow \varphi_{\{i,j\}}^{\{i,i\}}}
	\!\!=\!\! \prod_{\substack{k=1 \\ k \neq j}}^{Q} \big(\pi_{tr1}^{p,\{i,k\}}
	\cdot  \pi_{tr2}^{p,\{k,i\}}\big)
	\cdot g^B_{r,(i-1)Q+i}
\end{equation}

\begin{equation}
	\small
	\begin{aligned}
		&	v{(0)}_{s_{(i-1)Q+i} \rightarrow \varphi_{\{i,j\}}^{\{i,i\}}}
		= \prod_{\substack{k=1 \\ k \neq j}}^{Q} \bigg(\left(1 - \pi_{tr1}^{p,\{i,k\}}\right)\\&
		\cdot  \left(1 -  \pi_{tr2}^{p,\{k,i\}}\right)\bigg)
		\cdot \left(1 - g^B_{r,(i-1)Q+i}\right).
	\end{aligned}
\end{equation}
 $ \pi_{tr2}^{p,\{k,i\}}$ represents the probablity of $s_{(i-1)Q+i}=1$ in the message passing over the path $\varphi_{\{k,i\}}^{\{i,i\}}\rightarrow s_{(i-1)Q+i}$, whose detailed expression will be specified later.

5) Message Passing Over the Path $ \varphi_{\{i,j\}}^{\{i,i\}}\rightarrow s_{(i-1)Q+j}$: We denote the message passing over this path as $\nu_{\varphi_{\{i,j\}}^{\{i,i\}}\rightarrow s_{(i-1)Q+j}}$, which is given by
\begin{equation}\label{eq:0901_2}
	\small
	\begin{aligned}
		&\nu_{\varphi_{\{i,j\}}^{\{i,i\}}\rightarrow s_{(i-1)Q+j}}\\&\propto\!\!\! \sum_{s_{(i-1)Q+i}}\!\!\!\nu_{s_{(i-1)Q+i}\rightarrow\varphi_{\{i,j\}}^{\{i,i\}}}\times\varphi(s_{(i-1)Q+j},s_{(i-1)Q+i})\\&\propto\pi^{n,\{i,j\}}_{tp1}(\delta(s_{(i-1)Q+j}-1))+(1-\pi^{n,\{i,j\}}_{tp1})\delta(s_{(i-1)Q+j}),
	\end{aligned}
\end{equation}
where
\begin{equation}\label{eq:0901_3}
	\small
	\begin{aligned}
		\pi^{n,\{i,j\}}_{tp1}\triangleq\frac{\pi^{n,\{i,j\}}_{tr1}\cdot e^{\omega}}{1+\pi^{n,\{i,j\}}_{tr1}\cdot e^{\omega}},
	\end{aligned}
\end{equation}
represents the probability of $s_{(i-1)Q+j}=1$ in the message passing over the path $\varphi_{\{i,j\}}^{\{i,i\}}\rightarrow s_{(i-1)Q+j}$.

6) Message Passing Over the Path $s_{(i-1)Q+j}\rightarrow \varphi_{\{i,j\}}^{\{j,j\}}$: We denote the message passing over this path as $\nu_{s_{(i-1)Q+j}\rightarrow \varphi_{\{i,j\}}^{\{j,j\}}}$, which is given by
\begin{equation}\label{eq:0829_5}
	\small
	\begin{aligned}
		&	\nu_{s_{(i-1)Q+j}\rightarrow \varphi_{\{i,j\}}^{\{j,j\}}} \\&\propto g^B_{r,(i-1)Q+j}\times \!\!\nu_{ \varphi_{\{i,j\}}^{\{i,i\}}\rightarrow s_{(i-1)Q+j}}\\&\propto
		\pi^{p,\{i,j\}}_{tp2} \delta(s_{(i-1)+j}-1)+(1-	\pi^{p,\{i,j\}}_{tp2} )\delta(s_{(i-1)+j}),
	\end{aligned}
\end{equation}
where 
\begin{equation}\label{eq:0829_6}
	\small
\begin{aligned}
	\pi^{p,\{i,j\}}_{tp2} \triangleq 
\frac{\pi^{n,\{i,j\}}_{tp1}\,\pi^B_{(i-1)Q+j}}
{\pi^{n,\{i,j\}}_{tp1}\,\pi^B_{(i-1)Q+j} + \big(1-\pi^{n,\{i,j\}}_{tp1}\big)\big(1-\pi^B_{(i-1)Q+j}\big)},
\end{aligned}
\end{equation}
represents the probability of $s_{(i-1)Q+j}=1$ in the message passing over the path $s_{(i-1)Q+j}\rightarrow \varphi_{\{i,j\}}^{\{j,j\}}$.

7) Message Passing Over the Path $\varphi_{\{i,j\}}^{\{j,j\}}\rightarrow s_{(j-1)Q+j}$: We denote the message passing over this path as $\nu_{\varphi_{\{i,j\}}^{\{j,j\}}\rightarrow s_{(j-1)Q+j}}$, which is given by
\begin{equation}\label{eq:0829_2}
	\small
	\begin{aligned}
		&\nu_{\varphi_{\{i,j\}}^{\{j,j\}}\rightarrow s_{(j-1)Q+j}} \\&\propto	\sum_{s_{(i-1)Q+j}}\!\!\nu_{ s_{(i-1)Q+j}\rightarrow\varphi_{\{i,j\}}^{\{j,j\}}} \times\varphi(s_{(i-1)Q+j},s_{(j-1)Q+j})\\&\propto
		\pi^{p,\{i,j\}}_{tr2}\delta(s_{(j-1)+j}-1)+(1-	\pi^{p,\{i,j\}}_{tr2})\delta(s_{(j-1)+j}),
	\end{aligned}
\end{equation}
where
\begin{equation}
	\small
	\pi^{p,\{i,j\}}_{tr2} \triangleq
	\frac{\pi^{p,\{i,j\}}_{tp2} e^{\omega} + \left(1-\pi^{p,\{i,j\}}_{tp2}\right)}
	{\pi^{p,\{i,j\}}_{tp2} e^{\omega} + \pi^{p,\{i,j\}}_{tp2} + 2\left(1-\pi^{p,\{i,j\}}_{tp2}\right)}.
\end{equation}

8) Message Passing Over the Path $s_{(j-1)Q+j}\rightarrow\varphi_{\{i,j\}}^{\{j,j\}}$: We denote the message passing over this path as $\nu_{ s_{(j-1)Q+j}\rightarrow\varphi_{\{i,j\}}^{\{j,j\}}}$, which is given by
\begin{equation}\label{eq:0829_8}
	\small
	\begin{aligned}
		&\nu_{s_{(j-1)Q+j}\rightarrow\varphi_{\{i,j\}}^{\{j,j\}}} \\&\propto\!\!\!
		\prod_{k=1,k\neq i}^{Q}\!\! \nu_{\varphi_{\{k,j\}}^{\{j,j\}}\rightarrow s_{(j-1)Q+j}} \times  \nu_{\varphi_{\{j,k\}}^{\{j,j\}}\rightarrow s_{(j-1)Q+j}} \times g^B_{r,(j-1)Q+j}\\&\propto \pi^{n,\{i,j\}}_{tr2} \delta(s_{(j-1)+j}-1)+(1-	\pi^{n,\{i,j\}}_{tr2} )\delta(s_{(j-1)+j}),
	\end{aligned}
\end{equation}
where
\begin{equation}
	\small
	\pi^{n,\{i,j\}}_{tr2} \triangleq
	\frac{v(1)_{s_{(j-1)Q+j}\rightarrow\varphi_{\{i,j\}}^{\{j,j\}}}}
	{v(1)_{s_{(j-1)Q+j}\rightarrow\varphi_{\{i,j\}}^{\{j,j\}}}+v(0)_{s_{(j-1)Q+j}\rightarrow\varphi_{\{i,j\}}^{\{j,j\}}}},
\end{equation}
represents the probability of $s_{(j-1)Q+j}=1$ in the message passing over the path $s_{(j-1)Q+j}\rightarrow\varphi_{\{i,j\}}^{\{j,j\}}$, and we denote that
\begin{equation}
	\small
	v{(1)}_{s_{(j-1)Q+j}\rightarrow\varphi_{\{i,j\}}^{\{j,j\}}}
	= \prod_{\substack{k=1 \\ k \neq i}}^{Q} \pi_{tr1}^{p,\{k,j\}}
	\cdot  \pi_{tr2}^{p,\{j,k\}}
	\cdot g^B_{r,(i-1)Q+i}
\end{equation}

\begin{equation}
	\small
	\begin{aligned}
		&	v{(0)}_{s_{(i-1)Q+i} \rightarrow \varphi_{\{i,j\}}^{\{i,i\}}}
		= \prod_{\substack{k=1 \\ k \neq i}}^{Q} \bigg(\left(1 - \pi_{tr1}^{p,\{k,j\}}\right)\\&
		\cdot \left(1 -   \pi_{tr2}^{p,\{j,k\}}\right)\bigg)
		\cdot \left(1 - g^B_{r,(i-1)Q+i}\right).
	\end{aligned}
\end{equation}

9) Message Passing Over the Path $ \varphi_{\{i,j\}}^{\{j,j\}}\rightarrow s_{(i-1)Q+j}$: We denote the message passing over this path as $\nu_{\varphi_{\{i,j\}}^{\{j,j\}}\rightarrow s_{(i-1)Q+j}}$, which is given by
\begin{equation}\label{eq:0901_5}
	\small
	\begin{aligned}
		&\nu_{\varphi_{\{i,j\}}^{\{j,j\}}\rightarrow s_{(i-1)Q+j}}\\&\propto\!\!\! \sum_{s_{(j-1)Q+j}}\!\!\!\nu_{s_{(j-1)Q+j}\rightarrow\varphi_{\{i,j\}}^{\{j,j\}}}\times\varphi(s_{(i-1)Q+j},s_{(j-1)Q+j})\\&\propto\pi^{n,\{i,j\}}_{tp2}(\delta(s_{(i-1)Q+j}-1))+(1-\pi^{n,\{i,j\}}_{tp2})\delta(s_{(i-1)Q+j}),
	\end{aligned}
\end{equation}
where
\begin{equation}\label{eq:0901_6}
	\small
	\begin{aligned}
		\pi^{n,\{i,j\}}_{tp2}\triangleq\frac{\pi^{n,\{i,j\}}_{tr2}\cdot e^{\omega}}{1+\pi^{n,\{i,j\}}_{tr2}\cdot e^{\omega}}.
	\end{aligned}
\end{equation}

 }


\bibliographystyle{IEEEtran} 
\bibliography{ref2.bib} 
\end{document}